\documentclass[11pt]{article}

\usepackage[margin=1in]{geometry}
\usepackage{amsmath,amssymb,amsthm}
\usepackage{graphicx}
\usepackage{hyperref}
\usepackage{xcolor}
\usepackage{bm,mathrsfs}
\usepackage{caption,subcaption}
\usepackage{booktabs,float}
\usepackage{titlesec,needspace,setspace}

\definecolor{darkgreen}{rgb}{0.0, 0.5, 0.0}

\title{Engineering classical waves with quantized energy spectra in periodic media}

\author{
Arnaud Lazarus$^{1,2}$, Georgi Gary Rozenman$^{1}$ and John W. M. Bush$^{1}$\\[1ex]
{\small $^{1}$Department of Mathematics, Massachusetts Institute of Technology, Cambridge, MA, USA}\\
{\small $^{2}$Institut Jean Le Rond d'Alembert, CNRS UMR7190, Sorbonne Universit\'e, Paris, France}\\
{\small \texttt{arnaud.lazarus@sorbonne-universite.fr}}
}

\date{}

\begin{document}

\maketitle

\begin{abstract}
Field quantization is a central feature of modern physics, that underpins the concept of photons and forms the foundation of quantum electrodynamics as well as much of solid-state theory. Classical linear wave equations are not generally expected to reproduce the quantization arising in quantum systems without introducing additional ingredients such as ad hoc nonlinear constraints, resonant particle-wave couplings or stochastic background fields. Here, we show that appropriately engineered linear wave media can recover fundamental features evocative of energy quantization in quantum mechanics. The key is to tailor periodic media in which wave propagation is strongly suppressed, except over a discrete set of narrow pass bands. In this regime, stationary wave solutions exhibit discrete energy and frequency spectra analogous to those arising in quantum mechanics despite the underlying dynamics remaining linear. Owing to the universality of the proposed mechanism, these effects may be realized experimentally using mechanical, electrical, or electromagnetic waves in appropriately designed periodic media. This work opens new avenues for designing metamaterials that enable control over discrete wave states while strengthening the conceptual bridge between classical and quantum wave physics.
\end{abstract}


\section{Introduction}
\label{sec:introduction}

At the end of the nineteenth century, two seemingly opposing descriptions of light, originating from Newton and Huygens, were debated~\cite{shapiro1989huygens}; should light be regarded as a stream of particles or as a wave? The wave picture eventually prevailed, largely owing to Young's celebrated double-slit experiments~\cite{young1804bakerian} and to Maxwell's equations, which described light as an electromagnetic wave propagating in a vacuum~\cite{jackson1975electrodynamics}. This classical picture was challenged at the beginning of the twentieth century. In 1900, Planck rationalized the black-body radiation spectrum by introducing the notion of energy quantization~\cite{planck1900theorie}, and in 1905 Einstein interpreted the photoelectric effect in terms of localized quanta of light~\cite{einstein1905erzeugung,arons1965einstein}, later called photons. The insights of Planck and Einstein, later developed within the framework of quantum wave mechanics by Schr\"odinger~\cite{schrodinger1926undulatory}, established the framework for light-matter interactions. This evolution culminated in canonical quantization, wherein Maxwell's equations are promoted to operators~\cite{dirac1925fundamental} resulting in an electromagnetic field composed of harmonic modes with discrete energy levels~\cite{loudon2000quantum}\footnote{beyond semiclassical optics, in which the electromagnetic field remains classical while matter is described quantum mechanically~\cite{scully1997quantumoptics,hecht2017optics}}.

Field quantization in quantum mechanics introduced a deep conceptual divide between classical wave mechanics and the quantum description of wave phenomena. This divide is so pronounced that many counterintuitive features of the quantum realm are widely thought to be inaccessible to classical wave engineering. One notable exception is the recent development of topological mechanical metamaterials~\cite{susstrunk2015observation,nassar2020nonreciprocity, alu2025bright}, inspired by the quantum Hall effect, which realize topologically protected wave propagation in entirely classical systems~\cite{meeussen2016geared,yves2017crystalline,zheng2019observation}. However, even in this case the intrinsic quantization underlying quantum mechanics is not reproduced.

In this work, we explore an overlooked quantization mechanism in a classical wave system. Rather than invoking photons, localized particles, or quantized matter degrees of freedom, we show that a discrete, quantized frequency-energy spectrum can be reproduce from the linear dynamics of a classical wave propagating in an engineered periodic medium.
Specifically, we perform a Bloch-wave analysis of the generic one-dimensional d'Alembert wave equation with periodic coefficients~\cite{brillouin1946wave}. We focus on regions of parameter space in which the pass bands of waves propagating in an infinitely extended periodic medium become so narrow that the allowed propagation parameters effectively form a discrete set, numerically approaching a unique value per band. We then examine the influence of Dirichlet boundary conditions on the resulting frequency (or energy) spectrum of stationary waves, with the aim of comparing this classical framework to the familiar energy spectrum obtained from the canonical quantization of waves confined to a box.

We show that, in the narrow-pass-band regime, the propagating frequencies and the spatial structure of the corresponding stationary modes are governed by the eigenspectrum of a Hermitian operator mathematically equivalent to the Hamiltonian of a stationary Schr\"odinger equation. This correspondence makes it possible to exploit the well-established theory of quantum bound states to engineer periodic media with effectively discrete propagation spectra, given by simple analytical formulae rather than the typically continuous dispersion relations of standard periodic wave systems. When Dirichlet boundary conditions are imposed, the discrete spectrum of the resulting wave eigenvalue problem becomes numerically degenerate, and is simply the spectrum of a Schr\"odinger Hamiltonian, with a multiplicity given by the number of cells contained within the finite domain. Moreover, in the narrow-pass-band limit, a linear superposition principle emerges, whereby the spectrum of a composite medium formed by concatenating different sub-media is simply given by the union of their individual spectra. Leveraging this property, we identify the spatial modulation of the medium and the initial conditions required to reproduce an energy-frequency spectrum analogous to that of quantized electromagnetic waves in a one-dimensional cavity. Owing to the universality of the proposed mechanism for linear wave systems, this approach could enable new forms of quantum-inspired wave control across a wide range of physical platforms.

Section~\ref{quantum2} briefly reviews the canonical quantization of an electromagnetic field in a one-dimensional cavity, in order to familiarize the non-specialist reader with the quantum treatment of the relevant wave confinement problem. In Section~\ref{bandgaps}, we use Floquet-Bloch numerical analysis to investigate the narrow-pass-band limit for waves propagating in infinite periodic media and to highlight the resulting connections with quantum mechanics. In Section~\ref{Dirichlet} we characterize the influence of Dirichlet boundary conditions on the wave spectrum in this regime and propose a physical configuration that enables an analogy with the energy quantization of waves in a one-dimensional box. Our results are summarized in Section~\ref{conclu}.

\section{Quantization of an electric wave in a $1D$ empty cavity}
\label{quantum2}

We first show a brief example of the canonical quantization of an electromagnetic field in an empty cavity~\cite{loudon2000quantum}, the process by which modern quantum mechanics accounts for the energy quantization of a stationary wave field, be it a cavity mode of light (photon), a lattice vibrational mode (phonon), or a plasma oscillation mode (plasmon).

We consider a perfectly conducting optical cavity (e.g. an optical fiber) of length $L$ and uniform cross-sectional area $A$, aligned along the $x$-axis. Considering only the electric part of the electromagnetic field, assuming a single polarization of this electric field $\mathbf{E}(x,t) = E(x,t)\,\hat{y}$, and neglecting transverse variations, Maxwell's equations in vacuum reduce to~\cite{jackson1975electrodynamics}:
\begin{equation}
\label{eq:waveelec}
\frac{\partial^2 E(x,t)}{\partial x^2} - \epsilon_0\mu_0\frac{\partial^2 E(x,t)}{\partial t^2}=0, \qquad \text{where $E(0,t)=0$ and $E(L,t) = 0$}
\end{equation}
are Dirichlet boundary conditions corresponding to perfect electrical conductors. In Eq.(\ref{eq:waveelec}), $\epsilon_0$ and $\mu_0$ are the permittivity and permeability of free space, respectively. We recall that in a vacuum, $\epsilon_0\mu_0=1/c^2$ where $c$ is the speed of light. 

\begin{figure}[htbp]
\centering
	\includegraphics[width=1\columnwidth]{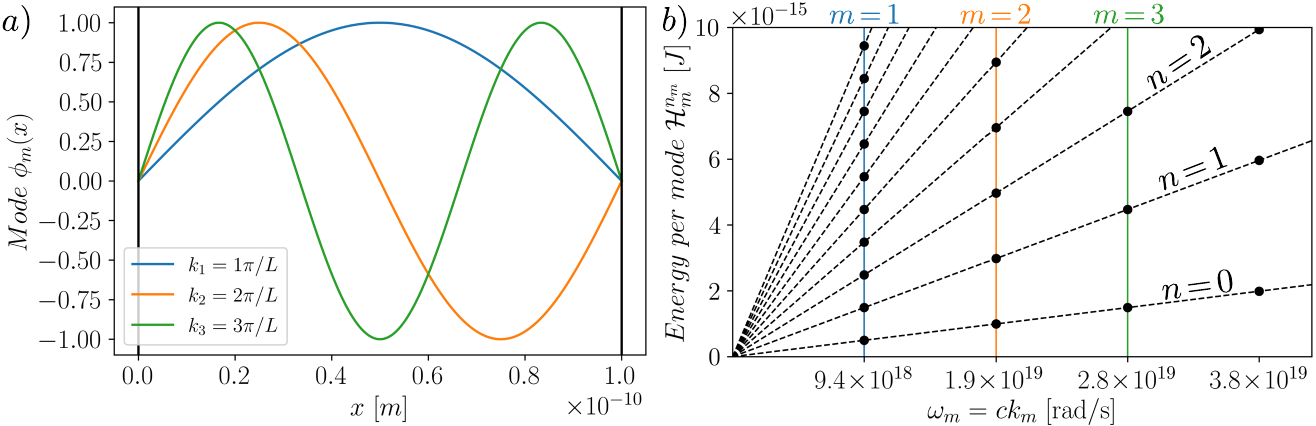}
	\caption{Energy field quantization of the electromagnetic wave in a $1D$ empty cavity. a) Allowed three first stationary ``modes'' $\phi_m(x)$ of the electric part of the wave for a box of length $L=1e^{-10}$ m. b) The total energy is proportional to each photon frequency $\omega_m=m\pi c/L$ and takes the form $\mathcal{E} = \sum_m\mathcal{H}^{n_m}_m = \sum_m (n_m + 1/2)\hbar \omega_m$.}
\label{fig:quantumener}
\end{figure}

We seek stationary solutions $E(x,t)$ of Eq.(\ref{eq:waveelec}) via separation of variables. Writing $E(x,t)=\phi(x)q(t)$ where $q(t) \propto e^{-i \omega t}$, one can rewrite Eq.(\ref{eq:waveelec}) in the form $\ddot{q}(t)/q(t)=c^2\phi''(x)/\phi(x)=-\omega^2$ where $\dot{(\,)}$ and $(\,)'$ denote derivatives with respect to time and space, respectively. One thus finds two separate equations in space and time
\begin{subequations}\label{Eqlinpodssdoreder}
\begin{align}
\phi''(x)+\frac{\omega^2}{c^2}\phi(x)=0
& \quad \text{with $\phi(0)=\phi(L)$=0,} \label{Eqlinpodssdorederspace}\\
\ddot{q}(t)+\omega^2q(t)=0
& \quad \text{with $q(0)=q_0$ and $\dot{q}(0)=v_0$.} \label{Eqlinpodssdoredertime}
\end{align}
\end{subequations}
Eq.(\ref{Eqlinpodssdorederspace}) implies that $\phi(x)$ can be expressed in the form $\phi(x)=A\cos(kx)+B\sin(kx)$ where $k=\omega/c$ is the dispersion relation relating the wavenumber $k$ and frequency $\omega$, and $A$ and $B$ are constants determined by the boundary conditions. Applying $\phi(0)=\phi(L)=0$, one finds that $A = 0$ and $B\sin(kL)=0$ so that non trivial electric waves only allow for a discrete set of wavenumber $
k_m
=m\pi/L$ (or $\omega_m=ck_m=m\pi c/L$) with $m=1,2,3,\dots,\infty$ associated to the eigenmodes $\phi_m(x)=B\sin(k_mx)$ as shown in Figure \ref{fig:quantumener}a.

In summary, the stationary electric field in the box of length $L$ can be written as a sum over its basis of orthogonal modes
\begin{equation}
E(x,t) = \sum_{m=1}^{\infty} q_m(t)\phi_m(x)= \sum_{m=1}^{\infty} q_m(t) \sin\left( \frac{m\pi x}{L} \right)
\label{Eq:modeprojectelec}
\end{equation}
where the modal amplitudes $q_m(t)$ are solutions of Eq.(\ref{Eqlinpodssdoredertime}) with $\omega_m=m\pi c/L$ and the initial conditions $q_m(0)=q^m_0$ and $\dot{q}(0)=v^m_0$.

If we assume the magnetic field is $B(x,0)=0$ at time $t=0$, the  total conserved energy of the oscillating electromagnetic field in the box of length $L$ and cross-sectional area $A$ can be defined in terms of its instantaneous electric field at time $t=0$ only
 \begin{equation}
\mathcal{E} = \frac{A}{2} \int_0^L \epsilon_0 |E(x,0)|^2 \, dx.
\label{Eq:classicalenerelec}
\end{equation}
Using the modal expansion given in Eq.(\ref{Eq:modeprojectelec}) at time $t=0$, $E(x,0)=\sum_mq_m^0\sin\left( \frac{m\pi x}{L} \right)$ and the orthogonality conditions $\int_0^L \sin\left(\frac{m\pi x}{L}\right)\sin\left(\frac{n\pi x}{L}\right) dx = \delta_{nm}L/2$ where $\delta_{nm}=1$ if $m=n$ or $0$ otherwise, one obtains the expression for the total energy 
\begin{equation}
\mathcal{E} = \frac{\epsilon_0AL}{4} \sum_{m=1}^{\infty}\left(q_0^m\right)^2.
\label{Eq:totalclassicalenerfinal}
\end{equation}
In Eq.(\ref{Eq:totalclassicalenerfinal}), the amplitude of the electric field at $t=0$, $q_0^m$, is undefined but usually chosen as a constant to respect the equipartition theorem, which dictates that each ``light'' mode inside the box has the same energy. 

Expression (\ref{Eq:totalclassicalenerfinal}) provides a good approximation in a wide range of physical situations, particularly at macroscopic volumes and high field intensities, such as radio-frequency wave propagation \cite{richards2008radiowave} or classical optics \cite{saleh2019fundamentals}.
It also underlies semiclassical descriptions of optical phenomena, in which the electromagnetic field is treated classically while its interaction with quantized matter accounts for discrete detection events \cite{griffiths2018introduction}. However, it does not capture the energy quantization of the electromagnetic field, $\mathcal{E} = \sum_{m=1}^{\infty}\left(n_m + \tfrac{1}{2}\right)\hbar \omega_m$, which becomes essential at low light intensities, where the field must be described in terms of discrete photons, as in the photoelectric effect or in the derivation of the black-body radiation spectrum \cite{schleich2015quantum}.

To predict the total energy in the box at low light intensities, one needs the linear algebra of quantum mechanics. First, one has no choice but to start from the complete form of the total conserved energy of the electromagnetic field in the box 
\begin{equation}
\mathcal{E}(t) = \frac{\epsilon_0}{2} \int_V |E(x,t)|^2 \, dV + \frac{1}{2\mu_0} \int_V |B(x,t)|^2 \, dV = \mathcal{E}.
\label{Eq:totalclassicalener}
\end{equation}
We know from Maxwell's equation that since $E(x,t) = \sum_{m=1}^{\infty} q_m(t) \sin\left( \frac{m\pi x}{L} \right)$ as shown in Eq.(\ref{Eq:modeprojectelec}), the magnetic field is of the form $B(x,t) = \sum_{m=1}^{\infty} b_m(t) \cos\left( \frac{m\pi x}{L} \right)$. The Amp\`ere-Maxwell law $\partial B(x,t)/\partial  x=(1/c^2)\partial E(x,t)/\partial  t$ implies that $b_m(t)=-L/(m\pi c^2)\dot{q}_m(t)=-\dot{q}_m(t)/(c\omega_m)$ since $\omega_m=m\pi c/L$. The total energy, Eq.(\ref{Eq:totalclassicalener}), then becomes
\begin{equation}
\mathcal{E} = \frac{\epsilon_0A}{2} \int_0^L |\sum_mq_m(t)\sin(\frac{m\pi x}{L})|^2 \, dx + \frac{A}{2\mu_0} \int_0^L |\sum_m\frac{\dot{q}_m(t)}{\omega_m c}\cos(\frac{m\pi x}{L})|^2 \, dx.
\label{Eq:totalclassicalenergeneralized}
\end{equation}
Using the orthogonality conditions $\int_0^L \cos\left(\frac{m\pi x}{L}\right)\cos\left(\frac{n\pi x}{L}\right) dx = \delta_{nm}L/2$, we may describe the system as a sum of decoupled oscillators whose total energy may be expressed as
\begin{equation}
\mathcal{E} = \sum_m\frac{\epsilon_0AL}{4} \left(q_m^2(t)+\frac{\dot{q}^2_m(t)}{\omega_m^2}\right)=\sum_m\mathcal{H} _m,
\label{Eq:totalclassicalfinal}
\end{equation}
where $\mathcal{H}_m$ is the so-called Hamiltonian of a given mode $m$ of a stationary oscillation of the electromagnetic wave in the box of length $L$. Introducing the analog mass $m_m=\epsilon_0AL/(2\omega_m^2)$, the analog stiffness $k_m=\epsilon_0AL/2$ and the analog momentum $p_m(t)=m_m\dot{q}_m(t)$, reveals that the Hamiltonian takes a well-known form analogous to that of a harmonic oscillator \cite{loudon2000quantum}
\begin{equation}
\mathcal{H}_m = \frac{1}{2}k_mq_m^2(t) + \frac{p_m^2}{2m_m}.
\label{Eq:totalclassicalfinal}
\end{equation}

From there, one needs to pass from classical to quantum description, the latter of which provides a natural formalism for obtaining quantization. We refer the reader to Supplementary Information S1 for a review of the full canonical quantization of light energy. Here, for the sake of clarity and brevity, we simply assume that if the harmonic oscillator Hamiltonian in Eq.(\ref{Eq:totalclassicalfinal}) is treated quantum mechanically rather than classically \cite{loudon2000quantum}, its energy only takes the discrete values
\begin{equation}
\mathcal E
= \sum_m \mathcal H_m^{n_m}
= \sum_{m=1}^{\infty} \hbar \omega_m \left( n_m + \tfrac{1}{2} \right),
\label{Eq:totalclassicalfinalquantum}
\end{equation}
where $\omega_m=ck_m=m\pi c/L$ is the frequency of the $m^{th}$ oscillatory mode, $\hbar$ is the reduced Planck constant and $n_m \in \mathbb{N}_0$ is the so-called occupation number and denotes the number of photons in mode $m$.


Two features of Eq.(\ref{Eq:totalclassicalfinalquantum}) are notable. First, the energy is actually proportional to the frequency of each photon mode in the box rather than the square of the field's amplitude as expected classically on the basis of Eq.(\ref{Eq:totalclassicalenerfinal}) or Eq.(\ref{Eq:totalclassicalfinal}). Second, each mode $m$ with frequency $\omega_m$ can only be measured with a discrete amount of energy $\mathcal{H}_m^{n_m}=\hbar\omega_m\left(n_m+\frac{1}{2}\right)$ and depends on the number of photons $n_m$. Fig.\ref{fig:quantumener}b shows the plot of $\mathcal{H}_m^{n_m}$  as a function of $\omega_m$ for a $1D$ box of length $L=1\times 10^{-10}$ m.
In quantum theory, $n_m=0$ for all modes $m$ is called the vacuum state, or ground state of a quantum field that is $\mathcal E=\sum_m\mathcal{H}_m^{0}= \sum_m\hbar\omega_m/2 \neq 0$. 
In the following, we show that it is possible to analogize Eq.(\ref{Eq:totalclassicalfinalquantum}) with a classical linear wave model. The approach is to consider the archetypal case of a wave propagating in a periodic medium, in a modulation parameter regime that has not previously been considered.

\begin{figure}[htbp]
\centering
	\includegraphics[width=1\columnwidth]{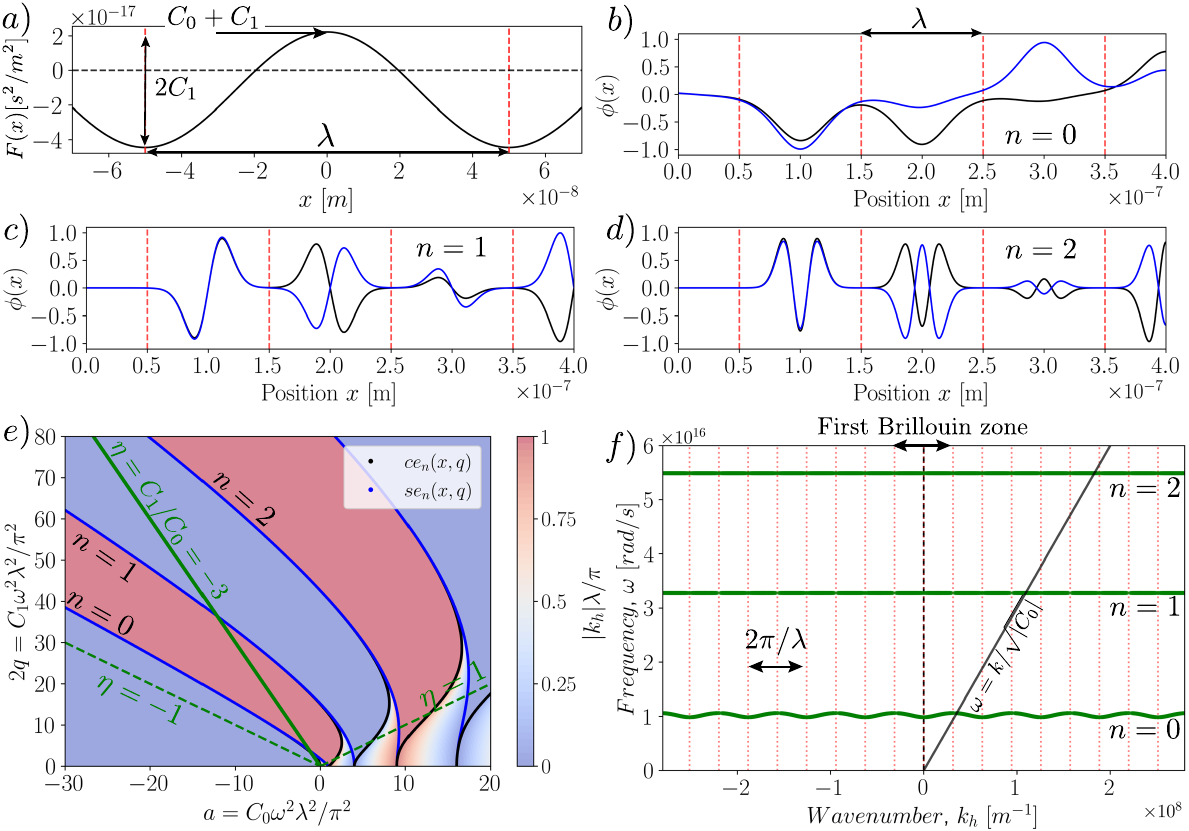}
	\caption{Waves in a medium for which the Helmholtz coefficient $F(x)=F(x+\lambda)$ is almost always negative only propagate in thin pass bands. a) Evolution of $F(x)=C_0+C_1\cos(2\pi x/\lambda)$ for $\lambda=1\times10^{-7}$ m, $C_0=-1/c^2$ and $\eta=C_1/C_0=-3$.  b) Snapshot of the spatial part $\phi(x)$ of two propagating waves $u(x,t)$ (normalized so that $max(\phi(x))=1$) whose frequency is randomly chosen in the first pass-band region $n=0$. c) Same as b) but for two waves in the band $n=1$. d) Same as b) but for the band $n=2$. e) Evolution of the Bloch wave dimensionless wavenumber $|k_h|\lambda/\pi$ belonging to the first Brillouin zone $-\pi/\lambda \leq k_h \leq \pi/\lambda$ in the $(a,2q)$ modulation space. The case $\eta=C_1/C_0=-3$ shown in a) is represented by a green line. f) Extended wave dispersion relation of a wave in a periodic medium with $F(x)$ given in a). The classic dispersion relation $\omega=k/\sqrt{|C_0|}$ for a medium with constant positive properties is indicated by the black line. We highlight the periodicity of the frequency in the wavenumber $k_h = k + h\pi/\lambda$ with $h \in \mathbb{Z}$ by plotting vertical dashed lines every $\pi/\lambda$.}
\label{fig:Floquetstyle}
\end{figure}

\section{Pass-band-induced quantization}
\label{bandgaps}

\subsection{Wave propagation in a harmonically varying one-dimensional medium}
\label{full}

The problem of band-gap formation in the propagation of waves in periodic continuous media is fundamental in classical physics~\cite{brillouin1946wave,richards2012analysis}. In the one-dimensional case, the stationary wave equation takes the Helmholtz form
\begin{equation}
\label{eq:Helmholtz}
\frac{\partial^2 u(x,t)}{\partial x^2} + \omega^2 F(x)u(x,t) = 0
\end{equation}
where $u(x,t)=\phi(x)e^{-i\omega t}$, and where the Helmholtz coefficient $F(x)$ is a $\lambda$-periodic function, with $\lambda$ the spatial period of the structured medium, assumed to be infinite in extent. A paradigmatic example is the harmonic modulation~\cite{brillouin1946wave}
\begin{equation}
\label{eq:Harmonicmod}
F(x)=C_0+C_1\cos(2\pi x/\lambda)
\end{equation}
shown in Fig.~\ref{fig:Floquetstyle}a, where $C_0$ and $C_1$ are two real coefficients with units of $1/c^2$. In the limit where $C_0>0$ and $C_1\rightarrow0$, Eq.~(\ref{eq:Helmholtz}) reduces to the classical wave equation in a homogeneous medium, with plane-wave solutions of the form $u(x,t)\propto e^{-i(\omega t \pm kx)}$ and linear dispersion relation $\omega=k/\sqrt{C_0}$. In this limit, $F(x)=C_0$ can be interpreted as the inverse squared wave speed of the medium. More generally, when $C_1\neq0$, the propagating solutions are no longer monochromatic plane waves but Bloch waves, as illustrated in Fig.~\ref{fig:Floquetstyle}b-d, whose dispersion relation exhibits band gaps as shown in Fig.~\ref{fig:Floquetstyle}f (see also Supplementary Info S2). In this regime, especially when $F(x)$ varies strongly or changes sign, $F(x)$ ceases to admit a simple interpretation in terms of a local wave speed. Rather, it acts as a spatial coefficient that controls the local character of the solutions, which are oscillatory when $F(x)>0$, and evanescent when $F(x)<0$. A locally negative $F(x)$ corresponds to a region where the underlying homogeneous state is linearly unstable, such that perturbations are amplified rather than propagated. Such sign-changing Helmholtz coefficients may arise in active or feedback-controlled media exhibiting locally non-passive propagation dynamics~\cite{PhysRevXZhu,fleury2014negative,wang2017active,veenstra2024non}.

To solve the Bloch problem given in Eqs.(\ref{eq:Helmholtz})-(\ref{eq:Harmonicmod}), Eq.~(\ref{eq:Helmholtz}) can be recast in the form of the Mathieu equation~\cite{whittaker1996course,kiorpelidis2024transient}
\begin{equation}
\label{eq:Mathieu}
\frac{d^2 \phi(\xi)}{d \xi^2} + \left[a + 2q\cos(2\xi))\right]\phi(\xi) = 0
\end{equation}
where $\xi=\pi x/\lambda$,  $a=\omega^2\lambda^2C_0/\pi^2$ and $2q=\omega^2\lambda^2C_1/\pi^2$. Now $a$ and $q$ are related by $a=2 q/\eta$ where $\eta=C_1/C_0$ represents the relative importance of the spatial modulation. 
Floquet theory dictates that $\phi(\xi)$ take the form
\begin{equation}
\label{eq:Floquetform}
\phi(\xi) = A\Psi(\xi)e^{s\xi} + B\overline{\Psi}(\xi)e^{-s\xi}\end{equation}
where $\Psi(\xi)$ is a complex $\pi$-periodic function with the same period as $F(\xi)$, $\overline{\Psi}(\xi)$ is its complex conjugate, $s$ is a complex number called the Floquet exponent, and $A$ and $B$ are arbitrary constants ~\cite{whittaker1996course}. 
In order to compute the dispersion relation, that is, the relation between $\omega$ and the wavenumber spectrum of $\phi(\xi)$ ~\cite{brillouin1946wave}, we choose the frequency $\omega$ as primary data and compute the Floquet form $\phi(\xi)$ given in Eq.~(\ref{eq:Floquetform}) with well-established numerical techniques from the literature~\cite{lazarus2015stability,bentvelsen2017modal}.

Depending on the modulation parameters $a$ and $q$, the Floquet exponent may be (i) $s=i\kappa$, purely imaginary, corresponding to bounded propagating Bloch waves with real wavenumber $\kappa$ and no exponential growth or decay, or (ii) $s=\sigma + i\kappa$, complex or real, corresponding to spatially attenuated (or amplified) waves whose amplitude varies exponentially along $\xi$. We recall that since $\Psi(\xi)$ is $\pi$-periodic, it can be expanded in a Fourier series $\Psi(\xi)=\sum_{h=-\infty}^{\infty}\Psi_he^{i2h}$ so that the actual wavenumber spectrum of the Floquet form $\phi(\xi)$ may be expressed as $s_h=\sigma +i\kappa_h$ where $\kappa_h = \kappa + 2h$ with $h \in \mathbb{Z}$. Figure~\ref{fig:Floquetstyle}e shows the evolution of the wavenumber $|\kappa_h|$ belonging to the so-called first Brillouin zone $-1 \leq |\kappa_h| \leq 1$, as a function of $a$ and $2q$. The plain-colored regions, referred to as stop bands in the language of photonics or solid-state physics \cite{meade2008photonic,kittel2018introduction}, correspond to parameters $(a,q)$ for which $\sigma \neq 0$, so that waves cannot propagate due to attenuation.  
The multicolored regions correspond to $\sigma=0$, where $s$ is purely imaginary; they are called pass bands, as they support wave propagation.

The two dashed lines $a=-2q$ (or $\eta=-1$) and $a=2q$ ($\eta=1$) shown in Fig.\ref{fig:Floquetstyle}e are drawn as guides. The typical cases of wave propagating in periodic media correspond to $(a,q)$ with $a$ positive and $0 < 2q < a$ where broad pass bands and narrow  stop bands occur (see Supplementary information S2). 
When $a<0$ and $a<-2q$, propagation is impossible because $F(x)$ is always negative and all waves are attenuated. A less typical case, highlighted by some researchers like Brillouin~\cite{brillouin1946wave} but never thoroughly investigated, corresponds to $(a,q)$ modulation parameters between the two dashed lines of Figure~\ref{fig:Floquetstyle}e for which $2q > |a|$, i.e. $F(x)$ is negative and positive over each wavelength of modulation $\lambda$. In this case, propagation of the wave is still possible although the width of the pass bands is small. 
In this manuscript, we investigate this narrow-pass-band limit.

To practically ensure the narrow-pass-band limit, we consider negative $C_0$ and positive $C_1$, i.e. negative $\eta=C_1/C_0=2q/a$ as in Figure~\ref{fig:Floquetstyle}a. Since $a=\omega^2\lambda^2C_0/\pi^2$ and $2q=\omega^2\lambda^2C_1/\pi^2$, increasing the frequency $\omega$ for a given medium is represented by a straight line with slope $\eta=2q/a$ running from the origin of $(a,2q)$, as illustrated by the green line of figure~\ref{fig:Floquetstyle}e. This line cuts successive narrow pass bands of pure imaginary Floquet exponent $s = i\kappa$. The resulting extended dispersion curve is given for $\eta=-3$ in figure~\ref{fig:Floquetstyle}f, showing the allowed $\omega$ as a function of the wavenumber spectrum of $\phi(x)$, $k_h=\pi/\lambda(\kappa+2h)$, for $F(x)=C_0+ C_1\cos(2\pi x/\lambda)$ with $\lambda=1\times 10^{-7}$ m and  $C_0=-1/c^2=-C_1/3$, as illustrated in figure~\ref{fig:Floquetstyle}a. 

We label the independent pass bands by an integer $n=0,1,2,\ldots$ as shown in figure~\ref{fig:Floquetstyle}f. As $\omega$ increases, the width of the pass bands decreases but always remains finite in theory. Because we consider an infinitely long system, each pass band contains an infinite number of Bloch waves $u(x,t)$ associated with Floquet multipliers $\rho=e^{ik\lambda}$ and $\bar{\rho}=e^{-ik\lambda}$, which lie on the unit circle in the complex plane so that $|\rho|=|\bar{\rho}|=1$. Figures~\ref{fig:Floquetstyle}b, c and d show the spatial part $\phi(x)$ of two typical bounded Bloch waves $u(x,t)\propto \phi(x)e^{-i\omega t}$ with randomly chosen $\rho$ in the first ($\rho = 0.529+0.848i$ and $\rho = 0.107+0.994i$), second ($\rho = -0.444+0.896i$ and $\rho = 0.211+0.977i$) and third ($\rho = -0.444+0.896i$ and $\rho = -0.535+0.844i$) pass-band region of figure~\ref{fig:Floquetstyle}f, respectively. We note that in the narrow-pass-band limit, a pass band $n=0,1,2,\ldots$ corresponds to a stationary wave with a specific signature: $\phi_n(x)$ has a typical shape with $(n+1)$ local maxima per wavelength of the periodic medium.

\begin{figure}[htbp]
\centering
	\includegraphics[width=1\columnwidth]{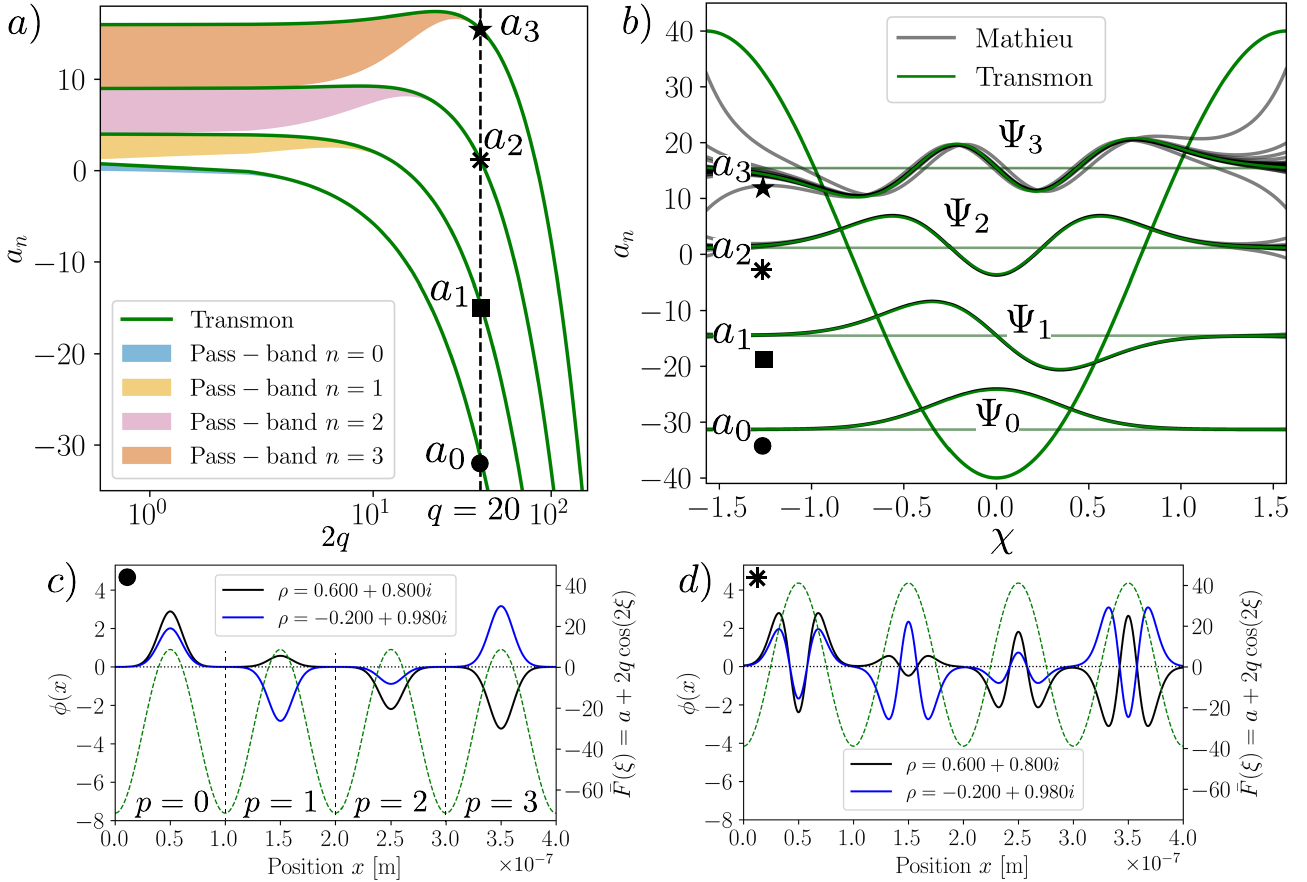}
	\caption{In the narrow-pass-band limit, the propagating waves can be described in terms of a stationary Schr\"odinger equation. a) First four pass-band regions of Figure \ref{fig:Floquetstyle}e as a function of $2q$. The eigenvalues $a_n$ of the Transmon equation Eq.(\ref{eq:Transmon}) are shown in green lines. b) Green lines show the typical representation of the first four wavefunctions $\Psi_n(\chi)$ of Eq.(\ref{eq:Transmon}) with a dimensionless potential $\bar{\mathcal{E}}=-2q\cos(2\xi)$ with $q=20$. The wavefunction amplitudes are plotted with a local origin corresponding to the eigenvalues $a_n$ of Eq.(\ref{eq:Transmon}) and the $\Psi_n(\chi)$ are normalized, $\int_{-\pi/2}^{\pi/2}\Psi_n(\chi)^2d\chi=1$, for visualization purposes. The stability regions $a_n$ of the Mathieu equation Eq.(\ref{eq:Mathieu}) for $q=20$ are reported with gray horizontal thin regions. The spatial part of two Bloch waves, $\phi(\xi)$, associated with Floquet multipliers $\rho=0.6+0.8i$ and $\rho=-0.2+0.98i$ in the first four $n^{th}$ stability regions are reported with gray lines. On each wavelength $p$, $\phi(x)$ have been divided by $\Re(C\rho^p)$ where $C$ is a complex constant that allows to fit the eigenfunctions of Eq.(\ref{eq:Transmon}). For the sake of clarity, we represent only the first 20 wavelengths of $\phi(\xi)$. c) Evolution of the spatial part of the propagating Bloch wave, $\phi(x)$, for $\rho=0.6+0.8i$ and $\rho=-0.2+0.98i$ as a function of $x$ for $\lambda=1\times 10^{-7}$ m, $q=20$ and $a$ in the pass-band region $n=0$. The associated dimensionless function $\bar{F}(\xi)=F(\xi)\omega^2\lambda^2/\pi^2=a+2q\cos(2\xi)$ is shown in dashed green lines. d) Same as c) but for $n=2$.}
\label{fig:reductiontransmon}
\end{figure}

\subsection{Spectral theory equivalence in the narrow pass-band limit}
\label{smallquantum}

Let's denote $\Psi_n(\xi)=\phi^{\rho=1}_n(\xi)$, with $n=0,1,2,\ldots$, the $\pi$-periodic solutions of Eq.(\ref{eq:Mathieu}) with $\rho=1$ which are located in the $(a,q)$ space at the border of each pass band. Because Eq.(\ref{eq:Mathieu}) is a Mathieu equation, $\phi^{\rho=1}_n(\xi)$ are simply even Mathieu functions $ce_{n}(\xi,q)$ for even integers $n$ and odd Mathieu functions $se_{n+1}(\xi,q)$ for odd integers $n$ as shown in Figure~\ref{fig:Floquetstyle}e. Consequently, the discrete set of parameters $a_n(q)$ leading to $\rho=1$ are the associated Mathieu characteristic values $a_n(q)$ and $b_n(q)$.

For a given dimensionless modulation parameter $q$, the solutions $\Psi_n(\xi)=\phi^{\rho=1}_n(\xi)$ and their associated $a_n$ are not only solutions of the Mathieu Eq.(\ref{eq:Mathieu}) but of the eigenvalue problem
\begin{equation}
\label{eq:Transmon}
\left[-\frac{d^2}{d\chi^2} -2q\cos(2\chi)\right]\Psi_n(\chi)=a_n\Psi_n(\chi),
\end{equation}
where $\chi \in [-\pi/2,\pi/2]$. Eq.(\ref{eq:Transmon}) is simply a rewritten Mathieu equation with a change of variables.  For a given $q$, the eigenvalues of Eq.(\ref{eq:Transmon}) are the  characteristic values $a_n(q)$ associated with $\pi$-periodic Mathieu functions. In this form over $[-\pi/2,\pi/2]$, Eq.(\ref{eq:Transmon}) is well-known in the physics community because it is the dimensionless form of the stationary Schr\"odinger equation of a transmon Qubit \cite{roth2022transmon}. Because $\phi^{\rho=1}_n(\xi)$ is $\pi$-periodic, computing the eigenfunction $\Psi_n(\chi)$ over $[-\pi/2,\pi/2]$ is sufficient to reconstruct the whole Floquet form $\phi^{\rho=1}_n(\xi)$ following
\begin{equation}
\label{eq:Mathieureconstruct}
\phi^{\rho=1}_n(\xi)=C\Psi_n(\chi) \qquad \text{for $\xi \in [p\pi,(p+1)\pi]$ with $\chi=\xi-p\pi-\pi/2$ and $p \in \mathbb{N}_0$}
\end{equation}
where $C$ is a fitting constant and $p \in \mathbb{N}_0$ is a non negative integer that prescribes the wavelength number.

Figure \ref{fig:reductiontransmon}a shows in green the evolution of the first four Mathieu characteristic values $a_n(q)$, eigenvalues of Eq.(\ref{eq:Transmon}), as a function of $2q$. An interesting observation can be made by plotting the first four pass-band regions of Figure \ref{fig:Floquetstyle}e in Figure \ref{fig:reductiontransmon}a. As expected, the eigenvalues $a_n(q)$ of Eq.(\ref{eq:Transmon}), are the boundaries of the pass bands. An often overlooked feature is that as $q$ increases and the narrow-pass-band assumption becomes reasonable, the whole pass-band region shrinks towards the eigenvalues $a_n(q)$ of the stationary Schr\"odinger equation Eq.(\ref{eq:Transmon}). Figure \ref{fig:reductiontransmon}b, which is a typical representation of the energy levels of quantum bound states of a transmon, shows that at $q=20$ for example, one gets already a good approximation of the location of the first four whole pass-band regions (horizontal gray regions) by computing the first eigenvalues of Eq.(\ref{eq:Transmon}) illustrated by horizontal green lines.

We now focus on the eigenfunctions $\Psi_n(\chi)$ of Eq.~(\ref{eq:Transmon}), still using $q=20$ as a numerical example. The first four eigenfunctions, $\Psi_0$, $\Psi_1$, $\Psi_2$ and $\Psi_3$, are shown with green lines in Fig.~\ref{fig:reductiontransmon}b. We recall that, up to a normalization factor, these functions are equivalent to the first four $\pi$-periodic Mathieu functions $\phi^{\rho=1}_n(\xi)$, i.e. the solutions of the Mathieu equation Eq.~(\ref{eq:Mathieu}) for $q=20$. As shown previously in the context of time-periodic dynamical systems~\cite{lazarus2019discrete,grandi2023new,lazarus2025meaningful,lazarus2025optimal}, when the $n^{\mathrm{th}}$ pass band becomes sufficiently narrow, the correspondence between $\phi^{\rho=1}_n(\xi)$ and $\Psi_n(\chi)$ given by Eq.~(\ref{eq:Mathieureconstruct}) extends to all waves $\phi^\rho_n(\xi)$ within the same band. For $|\rho|=1$, Eq.~(\ref{eq:Mathieureconstruct}) generalizes to
\begin{equation}
\label{eq:Mathieureconstruct2}
\phi^{\rho}_n(\xi)=\Re(C(\rho)\rho^p)\Psi_n(\chi)
\qquad
\text{for $\xi \in [p\pi,(p+1)\pi]$, $\chi=\xi-p\pi-\pi/2$, $p \in \mathbb{N}_0$}
\end{equation}
where $C(\rho)$ is a complex fitting constant that depends on $\rho$ and on the phase of the periodic medium. In other words, in the narrow-pass-band limit, the stationary parts $\phi(\xi)$ of the waves $u(\xi,t)=\Re(\phi(\xi)e^{-i\omega t})$ within the $n^{\mathrm{th}}$ pass band can all be expressed as sequences of scaled functions $\Psi_n(\chi)$, where $\Psi_n$ is the $n^{\mathrm{th}}$ eigenfunction of Eq.~(\ref{eq:Transmon}).

As a validation example, we now consider two numerical solutions $\phi(\xi)$ with Floquet multipliers $\rho = 0.6+0.8i$ and $\rho=-0.2+0.98i$, respectively, such that $|\rho|=1$, in the pass-band regions $n=0$, $1$, $2$ and $3$. The cases $n=0$ and $n=2$ are shown in Fig.~\ref{fig:reductiontransmon}c-d\footnote{The function $\bar{F}(\xi)=F(\xi)\omega^2\lambda^2/\pi^2=a+2q\cos(2\xi)$ has been shifted by $\lambda/2$ in space for visualization purposes. This shift is justified by the phase invariance of the Floquet multipliers $\rho$ in the Bloch problem.}. These functions $\phi(\xi)$ are divided wavelength-by-wavelength by $\Re(C(\rho)\rho^p)$ and reported with gray lines in Fig.~\ref{fig:reductiontransmon}b. The fitting constant $C(\rho)$ is chosen to best match the Transmon eigenfunctions $\Psi_n(\chi)$. The narrower the pass-band region, the more accurate the approximation given by Eq.~(\ref{eq:Mathieureconstruct2}), which explains why Eq.~(\ref{eq:Transmon}) starts failing to accurately approximate Eq.~(\ref{eq:Mathieu}) from $n=2$ onward at $q=20$. In the narrow-pass-band limit, the sole Transmon eigenvalue problem Eq.~(\ref{eq:Transmon}) accurately approximates the propagating waves $u(x,t)=\phi(x)e^{-i\omega t}$ of the original wave problem Eq.~(\ref{eq:Helmholtz}), since both $\phi(x)$ and $\omega$ can be reconstructed from the eigenfunctions $\Psi_n(\chi)$ and eigenvalues $a_n(q)$, respectively. The only missing information in the associated Schr\"odinger eigenproblem Eq.~(\ref{eq:Transmon}) is the value of the Floquet phase $\rho$, or equivalently, the finite width of the pass bands of the original wave problem.

\begin{figure}[htbp]
\centering
	\includegraphics[width=0.95\columnwidth]{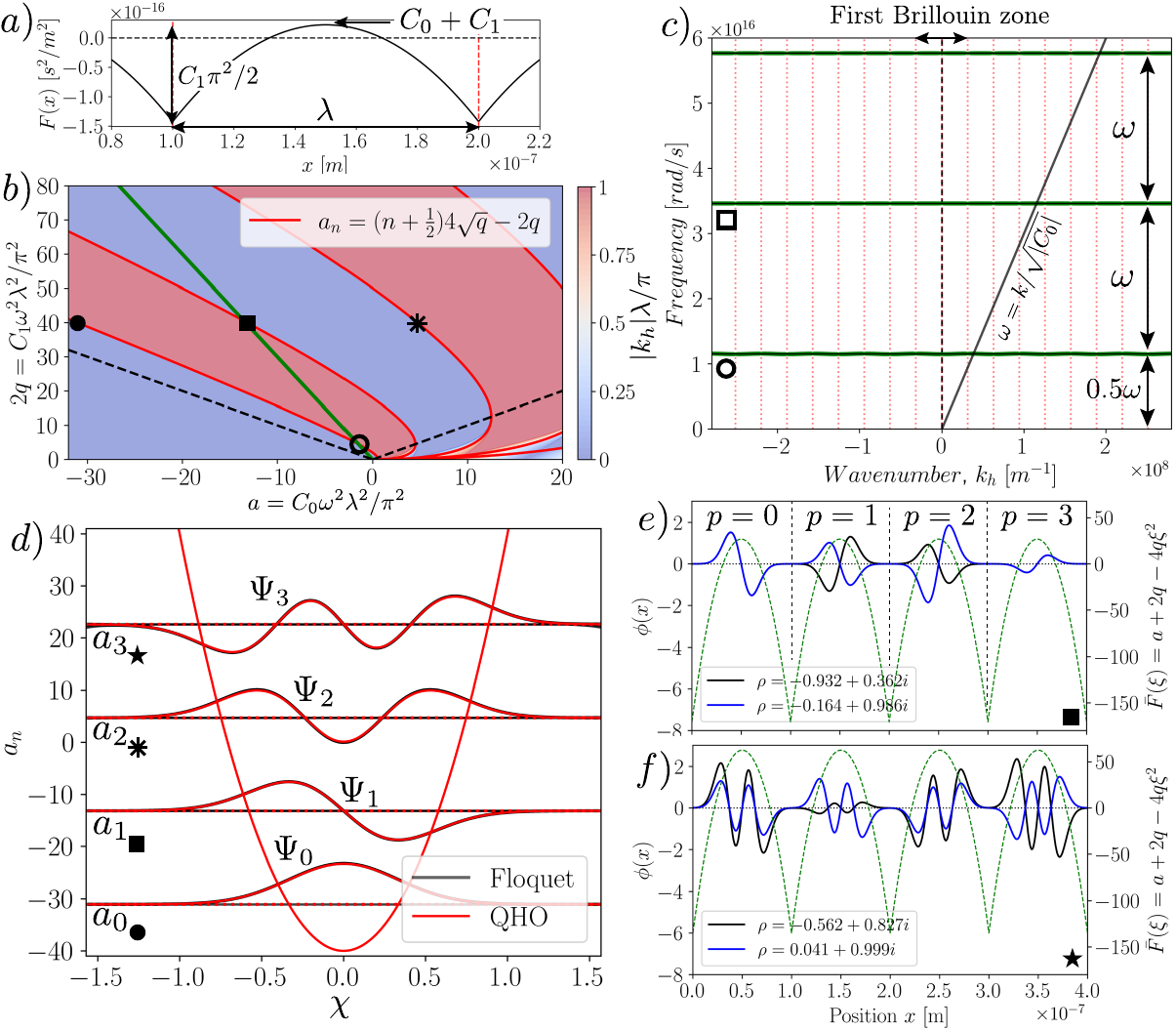}
	\caption{The medium, specifically the Helmholtz coefficient $F(x)=F(x+\lambda)$, can be designed so that propagating waves are governed by the mathematics of a Quantum Harmonic Oscillator (QHO). a) Form of $F(x)$ for $\lambda=1\times 10^{-7}$ m and $C_0=-1/c^2=-C_1/3$. b) Evolution of the Bloch wave dimensionless wavenumber $|k_h|\lambda/\pi$ belonging to the first Brillouin zone $-\pi/\lambda \leq k_h \leq \pi/\lambda$ in the $(a,2q)$ modulation space. The case $\eta=C_1/C_0=-3$ shown in a) is represented by a green line. c) Extended wave dispersion relation of the propagating waves for the periodic medium shown in a). The classic dispersion relation $\omega=k/\sqrt{|C_0|}$ is indicated by the black line for information. d) Typical representation of the first four wavefunctions of a QHO governed by Eq.(\ref{eq:QHO}) with a quadratic potential $4q\chi^2$ with $q=20$. The wavefunction amplitudes are plotted with a local origin corresponding to the eigenvalues $a_n=(n+1/2)4\sqrt{q}-2q$. The pass bands of the wave equation Eq.(\ref{eq:Hill}) for $q=20$ are the gray horizontal thin regions. The amplitude of the Floquet forms $\phi(\xi)$ associated with two different Floquet multipliers in the first four $n^{th}$ stability regions are reported with gray lines. On the twenty first wavelengths $p$, the Floquet forms have been divided by $\Re(C(\rho)\rho^p)$ where $C(\rho)$ is a complex constant that enables the fit to the eigenfunctions of Eq.(\ref{eq:QHO}). e) Evolution of the spatial part $\phi(x)$ of two  Bloch wave with $|\rho|=1$, $q=20$ and $a$ in the second pass-band region $n=1$. The associated $\bar{F}(\xi)$ is shown with green dashed lines. f) Same as e) but in the fourth pass-band region $n=2$.}
\label{fig:QHOFloquet}
\end{figure}

\subsection{Asymptotic equivalence to the Quantum Harmonic Oscillator}
\label{quantum}

We now reverse our perspective and investigate which modifications to the periodic medium are required to engineer a wave equation that becomes mathematically equivalent to the dimensionless stationary Schr\"odinger equation of the Quantum Harmonic Oscillator (QHO) in the narrow-pass-band limit. Recall that we consider the classical Helmholtz equation $\partial^2u(\xi,t)/\partial\xi^2+\omega^2F(\xi)u(\xi,t)=0$, which governs the stationary part of the $1D$ wave $u(\xi,t)=\Re(\phi(\xi)e^{-i\omega t})$, where $\xi=\pi x/\lambda$ is the dimensionless spatial coordinate and $F(\xi)=F(\xi+\pi)$ models the periodic properties of the medium. Substituting $u(\xi,t)$ into the Helmholtz equation yields an equation governing the spatial part
\begin{equation}
\label{eq:Hill}
\frac{d^2 \phi(\xi)}{d \xi^2} + \bar{F}(\xi)\phi(\xi) = 0
\end{equation}
where $\bar{F}(\xi)=F(\xi)\lambda^2\omega^2/\pi^2$ is the dimensionless function, which in the previous example was $\bar{F}(\xi)=a+2q\cos(2\xi)$. Using the second-order Taylor expansion of $\cos(2\xi)$ about $\xi=0$, namely $\cos(2\xi)\approx 1-2\xi^2$, in the Transmon eigenvalue problem Eq.~(\ref{eq:Transmon}) (see Supplemetary Information S4), we introduce the function $\bar{F}(\xi)=(a+2q)-4q\xi^2$. Using $a=C_0\lambda^2\omega^2/\pi^2$ and $2q=C_1\lambda^2\omega^2/\pi^2$, yields $F(\xi)$ of the form
\begin{equation}
\label{eq:FchiHQO}
F(\xi)=(C_0+C_1)-2C_1\chi^2 \quad \text{for $\xi \in [p\pi,(p+1)\pi]$, $\chi=\xi-p\pi-\pi/2$, $p \in \mathbb{N}_0$}.
\end{equation}
Figure~\ref{fig:QHOFloquet}a shows an example of this function $F(x)=F(\lambda \xi/\pi)$ for $\lambda=1\times 10^{-7}$ m and $C_0=-1/c^2=-C_1/3$. The evolution, as a function of $a$ and $2q$, of the Bloch wavenumber spectrum $|\kappa_h|=|\kappa + 2h|$ with $h \in \mathbb{Z}$ in the first Brillouin zone $-1 < \kappa_h < 1$ is shown in Figure~\ref{fig:QHOFloquet}b. Using $F(\xi)$ from Eq.(\ref{eq:FchiHQO}), the $\pi$-periodic functions $\Psi_n(\xi)=\phi^{\rho=1}_n(\xi)$ of Eq.(\ref{eq:Hill}) no longer satisfy the Transmon Eq.(\ref{eq:Transmon}) but the equivalent eigenvalue problem
\begin{equation}
\label{eq:QHO}
\left[-\frac{d^2}{d\chi^2}+4q\chi^2\right]\Psi_n(\chi)=(a_n+2q)\Psi_n(\chi)
\end{equation}
with $\chi \in [-\pi/2,\pi/2]$ and $\Psi_n(-\pi/2)=\Psi_n(\pi/2)$. In the narrow-pass-band limit, the functions $\phi^{\rho=1}_n(\xi)=\Psi_n(\xi)$ are localized within each wavelength, so that $\Psi_n(-\pi/2)=\Psi_n(\pi/2)\rightarrow 0$ is a reasonable approximation. Equation~(\ref{eq:QHO}) then becomes mathematically equivalent to the stationary Schr\"odinger equation of the QHO with quadratic potential $4q\chi^2$. The eigenvalues and eigenfunctions of the QHO are known in closed form~\cite{messiah1961quantum} and read
\begin{equation}
\label{eq:energyQHO}
a_n+2q=\left(n+\frac{1}{2}\right)4\sqrt{q}
\end{equation}
and
\begin{equation}
\label{eq:wavefunctionsHQO}
\Psi_n(\chi)=\frac{1}{\sqrt{2^nn!}}\left(\frac{2\sqrt{q}}{\pi}\right)^{1/4}e^{-\sqrt{q}\chi^2}H_n\left(\chi\left(2\sqrt{q}\right)^{1/2}\right)
\end{equation}
for $n \in \mathbb{N}_0$, where $H_n(z)=(-1)^ne^{z^2}\frac{d^n}{dz^n}\left(e^{-z^2}\right)$ are the physicists' Hermite polynomials. Equation~(\ref{eq:energyQHO}) is shown with red lines in Figure~\ref{fig:QHOFloquet}b. In the narrow-pass-band limit, Eq.(\ref{eq:energyQHO}) accurately predicts the location of $\phi^{\rho=1}_n(\xi)$ in the $(a,q)$ space, corresponding to the boundaries of the $n^{th}$ pass band. Since the pass bands are narrow, Eq.(\ref{eq:energyQHO}) also provides a good approximation of the location of the whole bands. 

Using the physical expressions of $a$ and $q$, Eq.(\ref{eq:energyQHO}) becomes
\begin{equation}
\label{eq:omegaQHO}
\omega=(n+\frac{1}{2})\frac{2\pi\sqrt{2C_1}}{\left(C_0+C_1\right)\lambda} \qquad \text{for $n \in \mathbb{N}_0$}
\end{equation}
which provides a closed-form expression for the frequencies that can propagate in a medium characterized by Eq.(\ref{eq:FchiHQO}) in the narrow-pass-band limit. Following the same procedure as in Figure \ref{fig:Floquetstyle}f, we plot in green the extended dispersion relation in Figure~\ref{fig:QHOFloquet}c, showing the allowed propagating frequencies $\omega$ as a function of the wavenumber spectrum of $\phi(x)$ for $\eta=2q/a=C_1/C_0=-3$ and $\lambda=1 \times 10^{-7}$ m. The discrete set of allowed frequency ranges, resulting from the extremely narrow pass bands as $|a|$ and $|q|$ increase, is well predicted by Eq.(\ref{eq:omegaQHO}), shown as thin horizontal black lines.

We now focus on the spatial structure of the propagating waves $\phi(\xi)$ within each pass band $n$, fixing for example $q=20$ as shown by the dark markers in Figure~\ref{fig:QHOFloquet}b. Each pass band contains an infinite number of propagating Bloch waves $u(\xi,t)=\phi(\xi)e^{-i\omega t}$ associated with Floquet multipliers satisfying $|\rho|=|e^{i(k+2h)\lambda}|=1$ for all $h \in \mathbb{Z}$. Figure~\ref{fig:QHOFloquet}d shows, for a quadratic potential $4q\chi^2$ with $q=20$, the eigenspectrum $(a_n,\Psi_n(\chi))$ of Eq.(\ref{eq:QHO}) for $n=0,\ldots,3$. Since $q=20$ corresponds to the narrow-pass-band limit for the first four bands, the eigenvalues $a_n$ coincide with the values approximating the location of the corresponding pass bands, as indicated by the dark markers in Figure~\ref{fig:QHOFloquet}b.

As in the previous section, the spatial part $\phi(\xi)$ of the propagating Bloch waves with $|\rho|=1$, solutions of Eq.(\ref{eq:Hill}), is related to the wavefunctions $\Psi_n(\chi)$ via
\begin{equation}
\label{eq:HQOreconstruct}
\phi^{\rho}_n(\xi)=\Re(C(\rho)\rho^p)\Psi_n(\chi) \quad \text{for $\xi \in [p\pi,(p+1)\pi]$, $\chi=\xi-p\pi-\pi/2$, $p \in \mathbb{N}_0$}.
\end{equation}
These functions $\phi(\xi)$, computed for $q=20$ and two values of $\rho$ located in the second and fourth pass bands (around $a_1$ and $a_3$), are shown in Figure~\ref{fig:QHOFloquet}e-f over four modulation wavelengths with $\lambda=1\times 10^{-7}$ m. Their associated $\bar{F}(\xi)$ is shown with dashed green lines. To illustrate Eq.(\ref{eq:HQOreconstruct}), the functions $\phi(\xi)$ are rescaled by $\Re(C(\rho)\rho^p)$ over the first 20 wavelengths and compared to the corresponding $\Psi_n(\chi)$ shown in Figure~\ref{fig:QHOFloquet}d. The agreement is excellent, confirming that the function $F(\xi)$ defined in Eq.(\ref{eq:FchiHQO}) provides an asymptotic equivalent of the QHO in the narrow-pass-band limit.

\section{Influence of Dirichlet boundary conditions
in the narrow-pass-band limit }
\label{Dirichlet}

To further pursue the analogy with the energy quantization described in Section~\ref{quantum2}, we now consider stationary waves confined within a finite box of length $L$ and subject to Dirichlet boundary conditions. We focus here on the narrow-pass-band limit arising when the Helmholtz coefficient $F(x)$ is negative almost everywhere, corresponding locally to evanescent wave behavior and to the absence of ordinary propagating-wave solutions. The more conventional case of a weak harmonic modulation of $F(x)$ about a positive value $C_0$ is treated separately in Supplementary Information S3.

\subsection{Stationary modes in a box}
\label{box_spectrum}

We proceed by studying the generic d'Alembert equation
 \begin{equation}
\label{eq:wavedalembert}
\frac{\partial^2 u(x,t)}{\partial x^2} - F(x)\frac{\partial^2 u(x,t)}{\partial t^2} = 0 \qquad \text{with $u(0,t) = u(L,t) = 0$}
\end{equation}
where the wave $u(x,t)$ is now prescribed in $x=0$ and $x=L$ by the Dirichlet boundary conditions $u(0,t) = u(L,t) = 0$ at any time $t$. To pursue the analogy with the field quantization of the electric field described in Section \ref{quantum2}, we choose $F(x)$ analogous to the QHO, specifically,
 \begin{equation}
\label{eq:periodmedia}
F(x)=F(x+\lambda)=(C_0+C_1)-2C_1\pi^2\bar{x}^2/\lambda^2
\end{equation}
where $\bar{x}=x-p\lambda-\lambda/2$, $x \in [p\lambda,(p+1)\lambda]$ and $p=0,1,2,\ldots,P$. To avoid complex boundary effects, we restrict ourselves to a box of size $L$ with a finite number $P$ of wavelengths $\lambda$ so that
 \begin{equation}
\label{eq:finitewavelengths}
L=P\lambda \qquad \text{with $P \in \mathbb{N}$.}
\end{equation}

Since Eq.(\ref{eq:wavedalembert}) is a linear PDE with a $\lambda$-periodic coefficient $F(x)$, Floquet-Bloch theory insures that we can seek  stationary solutions of the form $u(x,t)=\phi(x)q(t)$ with $\phi(x)=\phi(x+\lambda)$ and $q(t) \propto e^{-i \omega t}$. Substituting this expression into Eq.(\ref{eq:wavedalembert}), separation of variables yields,
\begin{equation}
\left\{
\begin{split}
\phi''(x)+\omega^2F(x)\phi(x)=0 & \quad \text{with $\phi(0)=\phi(L)$=0,} \\
\ddot{q}(t)+\omega^2q(t)=0 & \quad \text{with $q(0)=q_0$ and $\dot{q}(0)=v_0$}\\
\end{split}
\right.
\label{Eqlinperiod}
\end{equation}
where $q_0$ and $v_0$ are the initial conditions in modal space. As in Section \ref{quantum2}, Eq.(\ref{Eqlinperiod}) allows one to compute an orthonormal basis for $u(x,t)$ in the form $u(x,t)=\sum_m^{\infty}q_m(t)\phi_m(x)$ with $q_m(t)=q_0\cos(\omega_m t)+(v_0/\omega_m)\sin(\omega_m t)$ and $(\phi_m(x),\omega_m)$ the solution of an eigenvalue problem, here weighted by the function $F(x)=F(x+\lambda)$, that reads
 \begin{equation}
\label{eq:finiteoperator}
-\frac{d^2 \phi_m(x)}{dx^2}=F(x)\omega_m^2\phi_m(x) \quad \text{with $\phi_m(0)=\phi_m(L)=0$ $\forall m$}
\end{equation}
with $m \in \mathbb{N}$ the mode number.

When the box length is an integer multiple of the spatial period,
$L=P\lambda$ with $P\in\mathbb{N}$, the finite-domain eigenvalue
problem Eq.~(\ref{eq:finiteoperator}) can be interpreted as a restriction
of the infinite Bloch problem described in Section~\ref{bandgaps}.
For a frequency within a pass band, the infinite periodic
medium admits two Bloch solutions of the form
$\Psi(x)e^{ikx}$ and its complex conjugate
$\overline{\Psi}(x)e^{-ikx}$, where $\Psi(x+\lambda)=\Psi(x)$.
On the finite interval $[0,L]$, admissible eigenfunctions $\phi(x)$ are
linear combinations of these two counter-propagating waves.

Imposing the Dirichlet boundary conditions $\phi(0)=\phi(L)=0$
selects only specific values of the fundamental wavenumber $k$.
Since $L=P\lambda$ and $\Psi(x)$ is $\lambda$-periodic,
the boundary conditions restrict the Bloch phase $e^{ikL}$,
so that the continuous Bloch parameter becomes discrete:
\begin{equation}
k_\ell=\frac{\ell\pi}{L}=\frac{\ell\pi}{P\lambda},
\qquad \ell=1,\dots,P.
\label{eq:discretek}
\end{equation}

\begin{figure}[!t]
\centering
	\includegraphics[width=1\columnwidth]{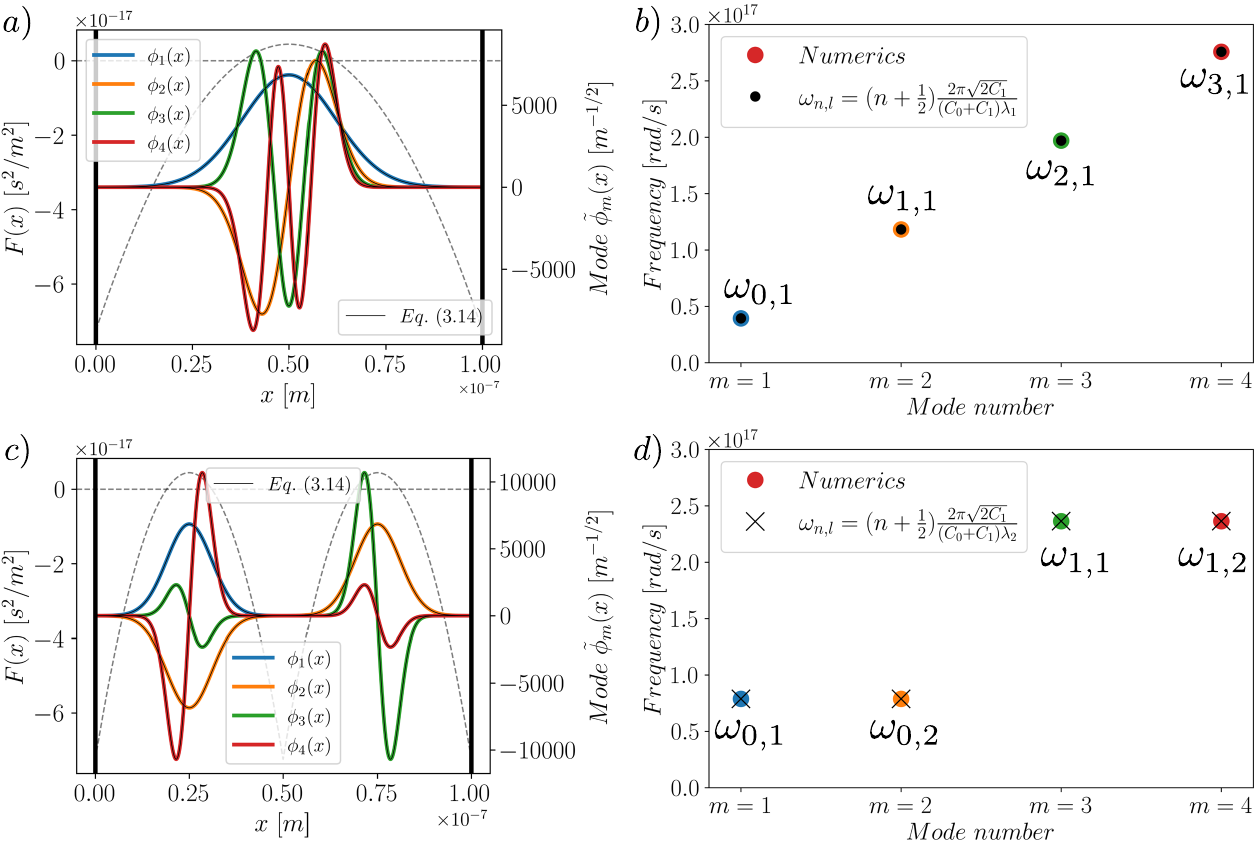}
	\caption{From propagating waves to standing modes. a) Evolution of the Helmholtz coefficient $F(x)$ and the first four stationary modes $\tilde{\phi}_m(x)$ of Eq.(\ref{eq:finiteoperator}) for $C_0=-1/c^2$, $C_1=1.4|C_0|$, $L=10^{-7}$ m and $\lambda = \lambda_1= L$. The modes $\tilde{\phi}_m(x)$ are normalized so that $\int_0^L\tilde{\phi}^2_m(x)dx=1$. The corresponding wavefunctions of the analogous QHO are shown in black lines. b) Oscillation frequencies $\omega_m$ of the modes in a). The analytical prediction from the QHO analogy is given in black markers. c-d) Same as a-b) but for an $F(x)$ with $\lambda = \lambda_2 = L/2$.}
\label{fig:HQOwavesinabox}
\end{figure}

Let $\omega_n(k)$ denote the dispersion relation of the $n$-th Bloch
band of the infinite periodic medium, $n=0,1,2,\dots$.
Evaluating these dispersion branches at the discrete wavenumbers
$k_\ell$ defined in Eq.~(\ref{eq:discretek}), modulo $2\pi/\lambda$,
yields the eigenfrequencies of the finite-domain problem Eq.~(\ref{eq:finiteoperator}).
More precisely, for each pass band $n$,
exactly $P$ eigenfrequencies are obtained in the reduced first Brillouin zone
$[0,\pi/\lambda]$, which read, for $\ell=1,\dots,P$:
\begin{equation}
\omega_{n,\ell} = \omega_n(k_\ell) \text{ for even $n$} 
\qquad 
\omega_{n,\ell} = \omega_n(k_{\ell-1}) \text{ for odd $n$.}
\label{discretewavi}
\end{equation}
Therefore, the spectrum of Eq.~(\ref{eq:finiteoperator}) consists of the
collection of $\omega_{n,\ell}$ which, when ordered increasingly, produces the
sequence of eigenfrequencies $\{\omega_m\}_{m\ge1}$ of the finite-domain
problem.

Figure \ref{fig:HQOwavesinabox} shows the first four $(\phi_m(x),\omega_m)$ of
Eq.~(\ref{eq:finiteoperator}) for $C_0=-1/c^2$, $C_1=1.4|C_0|$ and $L=1\times 10^{-7}$ m.
By setting $\eta=C_1/C_0=-1.4$, we ensure that the modulation parameters are in
the narrow-pass-band limit as defined in Section~\ref{bandgaps}.
Since $P$ eigenfrequencies $\omega_m = \omega_{n,\ell}$ are found per passing
band, the spectrum of Eq.~(\ref{eq:finiteoperator}) becomes numerically
degenerate in the narrow-pass-band limit, and the eigenvalues $\omega_m$
appear with multiplicity $P$, as shown, for example, in
Figure~\ref{fig:HQOwavesinabox}c for $P=2$. Moreover, unlike the typical case
shown in Supplementary Information S3 where closed form solutions are usually impossible, the eigenvalues $\omega_m$ simply tend to
\[
\omega_{n,\ell} =(n+1/2)\frac{2\pi\sqrt{2C_1}}{(C_0+C_1)\lambda},
\]
which correspond to the location of the pass bands of the associated
infinite Bloch problem given in Eq.~(\ref{eq:omegaQHO}) when $F(x)$ is chosen in
the QHO-analogous form of Eq.~(\ref{eq:periodmedia}). The eigenfunctions
$\phi_{m}(x)$ of Eq.~(\ref{eq:finiteoperator}) are also analytically known
since we have seen in Section~\ref{bandgaps} that, on each wavelength, they are
the solutions $\Psi_m(\chi)$ of the dimensionless Schr\"odinger equation of
a QHO
\begin{equation}
\label{eq:QHObis}
\left[-\frac{d^2}{d\chi^2}+4q_m\chi^2\right]\Psi_m(\chi)=(a_m+2q_m)\Psi_m(\chi)
\end{equation}
with $\Psi_m(-\pi/2)=\Psi_m(\pi/2)=0$, $q_m=C_1\omega_m^2\lambda^2/(2\pi^2)$,
$a_m=C_0\omega_m^2\lambda^2/\pi^2$ and $\chi=\bar{x}\pi/\lambda$.
Both Eq.~(\ref{eq:finiteoperator}) and Eq.~(\ref{eq:QHObis}) being linear
eigenvalue problems, their solutions are defined up to a normalization factor.
The dimensionless modes $\phi_m(x)$ are normalized by $\sqrt{\lambda/\pi}$ so
that $\int_0^L\tilde{\phi}_m(x)^2dx=1$ is dimensionless.
The $\Psi_m$ are normalized to fit the $\phi_m$ as shown, for example, in
Figure~\ref{fig:HQOwavesinabox}a for $P=1$. For $P>1$, this fitting factor on
each wavelength from $\Psi_m(\chi)$ to $\phi_m(\bar{x})$ is related to the
Floquet multiplier $\rho_m$ selected by the boundary conditions
$\phi(0)=\phi(L)=0$ from the unbounded Bloch problem, following
\begin{equation}
\label{eq:HQOreconstruct2}
\phi_m(x)=\Re(C(\rho_m)\rho_m^p)\Psi_m(\bar{x}),
\quad 
\text{$\bar{x}=x-p\lambda_m-\lambda_m/2$, $x \in [p\lambda_m,(p+1)\lambda_m]$, $p = 0,\ldots,P-1$}
\end{equation}
where $\Psi_m(\bar{x})$ is given in Eq.~(\ref{eq:wavefunctionsHQO}) and
$C(\rho_m)$ is the fitting constant chosen to match the 
normalization condition $\int_0^L\tilde{\phi}_m^2(x)dx=1$. Eq.~(\ref{eq:discretek}) works well in the unbounded Bloch scenario but sometimes fails in the finite case because of an apparent sensitivity of $\rho_m$ to the frequency in the narrow-pass-band limit when the spectrum of Eq.~(\ref{eq:finiteoperator}) becomes numerically degenerate. This sensitivity, already
highlighted in \cite{lazarus2025optimal} for linear time-periodic systems,
will be investigated in further studies.

\begin{figure}[htbp]
\centering
	\includegraphics[width=1\columnwidth]{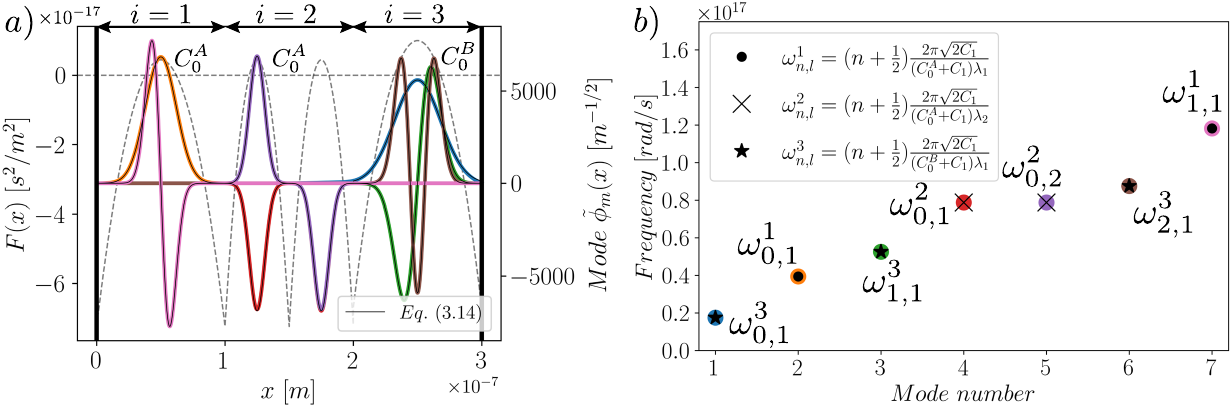}
	\caption{In the narrow-pass-band limit, decoupled spectrum from mode confinement enables tunable spectral engineering by medium concatenation. a) Evolution of a concatenated $F(x)$ and the first seven normalized stationary modes $\tilde{\phi}_m(x)$ of Eq.(\ref{eq:finiteoperator}). The medium in the box of length $L=3\times 10^{-7}$ m is such that $F(x)=F_1(x)$ $\forall x\in [0,L/3]$ with $C^1_0=C_0^A=-1/c^2$ and $\lambda=\lambda_1=L/3$, $F(x)=F_2(x)$ $\forall x\in [L/3,2L/3]$ with $C^2_0=C_0^A=-1/c^2$ and $\lambda=\lambda_2=L/6$, $F(x)=F_3(x)$ $\forall x\in [2L/3,L]$ with $C^3_0=C_0^B=0.5C_0^A$ and $\lambda=\lambda_1$. We fix $C^i_1=1.4/c^2$ for all $i$. The wavefunctions of the analogous QHO are given in black lines. b) Associated oscillation frequencies $\omega_m$. The analytical prediction from the QHO analogy for each $F_i(x)$ is shown with black markers.}
\label{fig:HQOwavesinaconcatenatebox}
\end{figure}

Another interesting property that emerges for waves propagating in a box with a periodically varying medium in the narrow-pass-band limit is illustrated in Figure~\ref{fig:HQOwavesinaconcatenatebox}. It is well known that if a function $F(x)$ is constructed by concatenating several segments, $F(x)=[F_1(x)\, F_2(x)\, \ldots]$, the eigenvalue problem in Eq.~(\ref{eq:finiteoperator}) will generally produce a spectrum that differs from that of the individual subproblems associated with each $F_i(x)$ (see, for example, Supplementary Information S5). Conversely, in the narrow-pass-band limit, the eigenvalues and eigenfunctions of Eq.~(\ref{eq:finiteoperator}) are simply the union of the spectra associated with each independent $F_i(x)$. Because the eigenmodes are compact and localized on each $F_i(x)$, there is no energy leakage between them, and stacking different $F_i(x)$ segments does not modify the spectrum. This is shown in Figure~\ref{fig:HQOwavesinaconcatenatebox}, where we concatenated the $F_i(x)$ of Figure~\ref{fig:HQOwavesinabox} as well as a third $F_i(x)$ with a different $C_0$. Each eigenmode $\phi_m(x)$ remains localized within its wavelength and, because each of them locally satisfies Eq.~(\ref{eq:QHObis}), the $\omega_m=\omega^i_{n_i,l_i}$ and the shape of the eigenfunctions $\phi_m(x)=\phi^i_{n_i,l_i}(x)$ are analytically predicted by the closed-form expressions of the bounded energies and wavefunctions of the dimensionless QHO given in Eq.~(\ref{eq:omegaQHO}) and Eq.~(\ref{eq:wavefunctionsHQO}). This property offers a unique design opportunity for metamaterials composed of mostly negative, periodic $F(x)$.

\subsection{Energy of the waves}
\label{box_energy}

In this section, we compute the total energy of the wave $u(x,t)=\sum_m\tilde{\phi}_m(x)\tilde{q}_m(t)$ under the standard initial conditions of modal energy equipartition, with the aim of drawing an analogy with photon field quantization. The energy expressions used are well known; their derivation is presented in Supplementary Information S5 for the interested reader.

The total energy of $u(x,t)$, a solution of the wave equation Eq.(\ref{eq:wavedalembert}), may be expressed as
\begin{equation}
\tilde{\mathcal{E}}
= \frac{1}{2}\left[ \int_0^L \left(\frac{\partial u}{\partial x}\right)^2 dx
     + \int_0^L F(x)\left(\frac{\partial u}{\partial t}\right)^2 dx \right].
\label{eq:energy_physical2}
\end{equation}
This quantity is conserved in time and is therefore fully determined by the initial conditions $u(x,0)$ imposed on the wave. Projecting Eq.(\ref{eq:wavedalembert}) onto its orthogonal normalized modal basis $\tilde{\phi}_m(x)$ computed with Eq.(\ref{eq:finiteoperator}) allows one to write the energy as
\begin{equation}
\tilde{\mathcal{E}}=\sum_m\left(\frac{1}{2}\tilde{k}_m\tilde{q}^2_m(t) + \frac{1}{2}\tilde{m}_m\dot{\tilde{q}}^2_m(t) \right) = \sum_m \frac{1}{2}\tilde{m}_m\left(\omega_m^2 \tilde{q}^2_m(t) + \dot{\tilde{q}}^2_m(t) \right) = \sum_m \tilde{\mathcal{H}}_m
\label{Eq:energymodal}
\end{equation}
where $\tilde{\mathcal{H}}_m$ is the so-called Hamiltonian of the $m^{th}$ mode of vibration,
\begin{equation}
\tilde{k}_m = \int_0^L \tilde{\phi}'_m(x)^2\,dx \qquad \text{and} \qquad \tilde{m}_m=\int_0^L F(x)\tilde{\phi}_m(x)^2\,dx
\label{Eq:weakformfrequency}
\end{equation}
are the effective stiffness and mass of mode $m$, and $\omega_m^2=\tilde{k}_m/\tilde{m}_m$.

We see from Eq.(\ref{Eq:energymodal})-(\ref{Eq:weakformfrequency}) that, in the case of a homogeneous medium where $F(x)$ has a constant positive value, the normalization $\int_0^L\tilde{\phi}_m(x)^2dx=1$ ensures $\tilde{m}_m$ is a constant for all $m$. Modal equipartition of energy $\tilde{\mathcal{H}}_m$ is then obtained by imposing the initial conditions
\begin{equation}
\tilde{q}_m(0) = \tilde{Q}_0/\omega_m \qquad \text{and} \qquad \dot{\tilde{q}}_m(0) = \tilde{V}_0 \qquad \text{for all $m$}
\label{Eq:inicond}
\end{equation}
where $\tilde{Q}_0$ and $\tilde{V}_0$ are two constants. When $F(x)=F(x+\lambda)$ varies periodically about a positive value, as in the typical example of Supplementary Information S5, $\tilde{m}_m=\int_0^LF(x)\tilde{\phi}_m(x)^2dx$ usually depends on $\tilde{\phi}_m(x)$ and the energy varies from one mode $m$ to another for this set of initial conditions. In the narrow-pass-band limit, we will demonstrate that mode confinement and quantum analogies lead to a $\tilde{m}_m$ independent of mode $m$ for a given amplitude of $F(x)$.

Let us consider a box of length $L$ whose medium consists of $M$ concatenated sub-media. Each sub-medium is characterized by an almost everywhere negative function $F_i(x)$, defined in Eq.~(\ref{eq:periodmedia}), with fixed amplitudes $C_0^i$, $C_1^i$, and containing $P_i$ wavelengths $\lambda_i$. Owing to the mode localization properties illustrated, for example, in Figure \ref{fig:HQOwavesinaconcatenatebox}, the initial conditions of Eq.(\ref{Eq:inicond}) can be reordered in the form $\tilde{q}_m(0)=\tilde{q}^i_{n_i,l_i}(0)=\tilde{Q}_0/\omega^i_{n_i,l_i}$ and $\dot{\tilde{q}}_m(0)=\dot{\tilde{q}}^i_{n_i,l_i}(0)=\tilde{V}_0$ for $i=1,\ldots,M$, $l_i=1,\ldots,P_i$ and where the band index $n_i$ takes values in a subset $\mathcal{N}_i$. The total energy, given in Eq.~(\ref{Eq:energymodal}), can then be expressed as a sum over the contributions of each sub-medium $F_i(x)$, each with multiplicity $P_i$, as
\begin{equation}
\tilde{\mathcal{E}} 
= \sum_{i=1}^{M} P_i \sum_{n_i \in \mathcal{N}_i} \tilde{\mathcal{H}}^i_{n_i}
= \sum_{i=1}^{M} P_i \sum_{n_i \in \mathcal{N}_i}  \frac{1}{2} \tilde{m}^i_{n_i} \left( \tilde{Q}_0^2 + \tilde{V}_0^2 \right),
\label{Eq:energymodalt0}
\end{equation}
where $\tilde{\mathcal{H}}^i_{n_i}$ and $\tilde{m}^i_{n_i}$ denote the Hamiltonian and effective modal mass associated with the $n_i$-th band mode of the $i$-th sub-medium, respectively.  Substituting $\tilde{m}^i_{n_i}$ from Eq.~(\ref{Eq:weakformfrequency}) and introducing the dimensionless local coordinate $\chi_i = \pi \bar{x}_i / \lambda_i$ within each wavelength $\lambda_i$, we obtain
\begin{equation}
\label{eq:Enerlimit}
\tilde{m}^i_{n_i} 
= \int_{-\pi/2}^{\pi/2} F_i(\chi_i)\,\phi^i_{n_i,l_i}(\chi_i)^2\,d\chi_i
= \int_{-\pi/2}^{\pi/2} \left[\left(C^i_0 + C^i_1\right) - 2C^i_1 \chi_i^2 \right] \Psi^i_{n_i}(\chi_i)^2\,d\chi_i,
\quad \forall\, l_i.
\end{equation}
Here, $\Psi^i_{n_i}(\chi_i)$ denotes the QHO wavefunction of order $n_i$ localized in the narrow-pass-band limit to the extent that $\int_{-\pi/2}^{\pi/2}\Psi^i_{n_i}(\chi_i)^2\,d\chi_i=1$ (see Eq.~(\ref{eq:wavefunctionsHQO})). Due to this mode localization and the normalization condition $\int_0^L \tilde{\phi}_m(x)^2 dx = 1$, the effective mass $\tilde{m}^i_{n_i}$ can be evaluated within a single representative cell of size $\lambda_i$. Furthermore, using the second moment of the QHO probability density in the scaled coordinate $\chi$, $\langle\chi^2\rangle_n=(n+1/2)/2\sqrt{q}$, we can write 
\begin{equation}
\label{eq:momentHQO}
\langle\chi_i^2\rangle_m=\langle\chi_i^2\rangle^{\ell_i}_{n_i}=\int_{-\infty}^{\infty}\chi_i^2\Psi^i_{n_i}(\chi)^2d\chi=\frac{n_i+1/2}{2\sqrt{q_i}}=\frac{a_i+2q_i}{8q_i}=\frac{C_0^i+C_1^i}{4C_1^i}
\end{equation}
where we have used $a/2q=C_0/C_1$ and $(n+1/2)/\sqrt{q}=(a+2q)/4q$ that we derived in Eq.(\ref{eq:energyQHO}) for the Bloch wave problem and that holds for $F_i(x)$ with a finite number of wavelengths $\lambda_i$. Provided the modes $\Psi_m(\chi)$ are well localized inside $[-\pi/2,\pi/2]$, as assumed by the narrow-pass-band limit, Eq.(\ref{eq:momentHQO}) holds on this interval and the effective mass $\tilde{m}^i_{n_i}$ finally becomes
 \begin{equation}
\label{eq:finalHm}
\tilde{m}^i_{n_i} = \left(C^i_0+C^i_1\right)\int_{-\pi/2}^{\pi/2}\Psi^i_{n_i}(\chi_i)^2d\chi_i-2C^i_1\langle\chi_i^2\rangle^{\ell_i}_{n_i} =\frac{C_0^i+C_1^i}{2}
\end{equation}
which depends solely on the amplitude of $F_i(x)$.

\begin{figure}[htbp]
\centering
	\includegraphics[width=1\columnwidth]{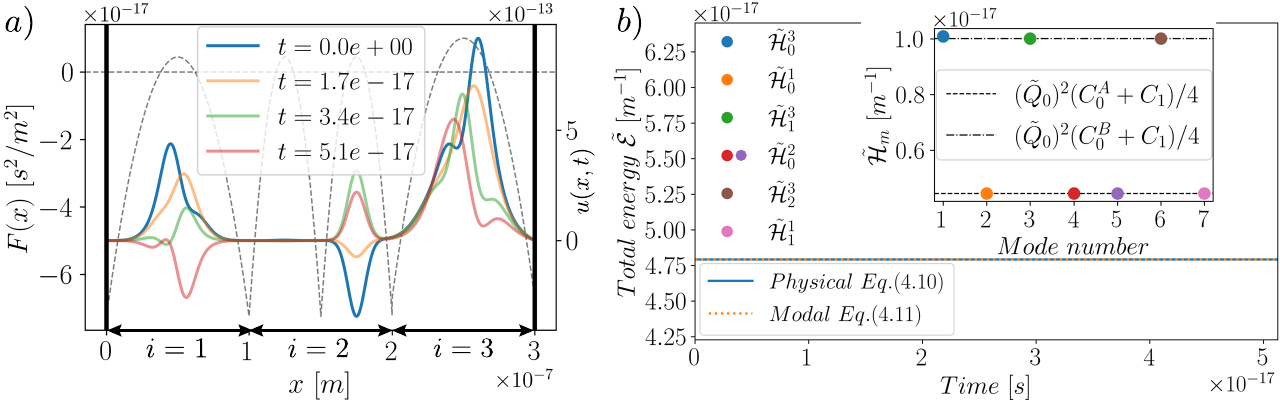}
	\caption{The QHO analogy allows one to recover mode-independent energy in the limit of a Helmholtz coefficient $F(x)$ that is almost everywhere negative. a) Four time steps showing the evolution $u(x,t)=\sum_{m=1}^7 \tilde{\phi}_m(x)\tilde{q}_m(t)$ in the box for the modal basis shown in Figure \ref{fig:HQOwavesinaconcatenatebox}. The initial conditions on each mode are $\tilde{q}_m(0)=\tilde{Q}_0/\omega_m$ with $\tilde{Q}_0=2$ $m^{1/2}s^{-1}$ and $\dot{\tilde{q}}_m(0)=0$ for all $m$. b) Total energy $\tilde{\mathcal{E}}$ in the box as a function of time computed in the physical (Eq.(\ref{eq:energy_physical2})) and modal space (Eq.(\ref{Eq:energymodal})).  Inset shows the energy decomposition in $\tilde{\mathcal{H}}_m$ following Eq.(\ref{Eq:energymodal}) or $\tilde{\mathcal{H}}^i_{n_i}$ following Eqs.(\ref{Eq:energymodalt0}) and (\ref{eq:finalHm}). }
\label{fig:HQOwavesinaboxenergy}
\end{figure}

Figure \ref{fig:HQOwavesinaboxenergy}a shows four time steps of the evolution $u(x,t)=\sum_m \tilde{\phi}_m(x)\tilde{q}_m(t)$ for the concatenated medium shown in Figure \ref{fig:HQOwavesinaconcatenatebox} and a modal basis $\tilde{\phi}_m(x)$ for $m=1,\ldots,7$. The initial condition is chosen to be $\tilde{q}_m(0)=\tilde{Q}_0/\omega_m$ with $\tilde{Q}_0=2$ $m^{1/2}s^{-1}$ and $\dot{\tilde{q}}_m(0)=0$ for the $N=7$ modes. The total energy $\tilde{\mathcal{E}}$ of the stationary wave $u(x,t)$ is given in Figure \ref{fig:HQOwavesinaboxenergy}b. As expected, the computation in physical or modal space, through Eq.(\ref{eq:energy_physical2}) or Eq.(\ref{Eq:energymodal}) respectively, leads to the same result that is a total energy $\tilde{\mathcal{E}}$ conserved with time. The inset of Figure \ref{fig:HQOwavesinaboxenergy}b shows the energy per mode $\tilde{\mathcal{H}}_m$ so that  $\tilde{\mathcal{E}}=\sum_1^7\mathcal{H}_m$. Since we are mostly in the narrow-pass-band limit for all the chosen modes (note the mode $m=1$ is barely localized in $F_3(x)$), the energy can also be decomposed into the sum of $\tilde{\mathcal{H}}^i_{n_i}$ according to Eq.(\ref{Eq:energymodalt0}) and (\ref{eq:finalHm}). Given $F(x)=[F_1(x),F_2(x),F_3(x)]$ described in Figure \ref{fig:HQOwavesinaconcatenatebox}, we can write $\tilde{\mathcal{E}}=1\times \sum_{n_1=0}^{1}\tilde{\mathcal{H}}^1_{n_1}+2\times \tilde{\mathcal{H}}^2_{0} + 1\times \sum_{n_3=0}^{2}\tilde{\mathcal{H}}^3_{n_3}$ with $\tilde{\mathcal{H}}^1_{n_i}=\tilde{\mathcal{H}}^2_{n_i}=\tilde{Q}_0^2(C_0^A+C_1)/4$ and $\tilde{\mathcal{H}}^3_{n_i}=\tilde{Q}_0^2(C_0^B+C_1)/4$ for all $n_i \in \mathbb{N}_0$.

\subsection{Numerical application to analogize the energy of photons}
\label{box_quantum}

Let's finish with an application to the particular case of Maxwell's equation in $1D$ as described in Section \ref{quantum2}. Here, instead of a classic vacuum with constant properties $F(x)=1/c^2$, we now consider electric waves oscillating in an exotic medium whose Helmholtz periodic coefficient $F(x)$ is in the narrow-pass-band limit. The differential equation governing the oscillation of electric field $E(x,t)$ in this medium is no longer Eq.(\ref{eq:waveelec}) but a more general equation
\begin{equation}
\label{eq:waveelecbis}
\frac{\partial^2 E(x,t)}{\partial x^2} - F(x)\frac{\partial^2 E(x,t)}{\partial t^2}=0 \qquad \text{where} \quad E(0,t)=E(L,t)=0.
\end{equation}
We recall from Section \ref{quantum2} that the classical expression of the total conserved energy of the oscillating electromagnetic field in the box of length $L$ and cross-sectional area $A$ can be defined in terms of its instantaneous electric field at time $t=0$ only
 \begin{equation}
\mathcal{E} = \frac{A}{2\mu_0} \int_0^L F(x) |E(x,0)|^2 \, dx
\label{Eq:classicalenerelecter}
\end{equation}
where $\mu_0$ is the vacuum permeability. Expanding the electric field in its modal basis, $E(x,t)=\sum_m \tilde{\phi}_m(x)\tilde{q}_m(t)$, we can transform Eq.(\ref{Eq:classicalenerelecter}) into modal space
 \begin{equation}
\mathcal{E} = \frac{A}{2\mu_0}\sum_m\tilde{q}_m^2(t)\int_0^LF(x)\tilde{\phi}_m
(x)^2dx=\sum_m\frac{A}{2\mu_0}\tilde{m}_m\tilde{q}_m^2(t)=\sum_m\mathcal{H}_m
\label{Eq:classicalenerelecmodal}
\end{equation}
where we recall $\tilde{m}_m=\int_0^L F(x)\tilde{\phi}_m(x)^2\,dx$ is the effective modal mass and $\mathcal{H}_m$ is the Hamiltonian contribution of each mode $m$, here in Joules.

\begin{figure}[htbp]
\centering
	\includegraphics[width=1\columnwidth]{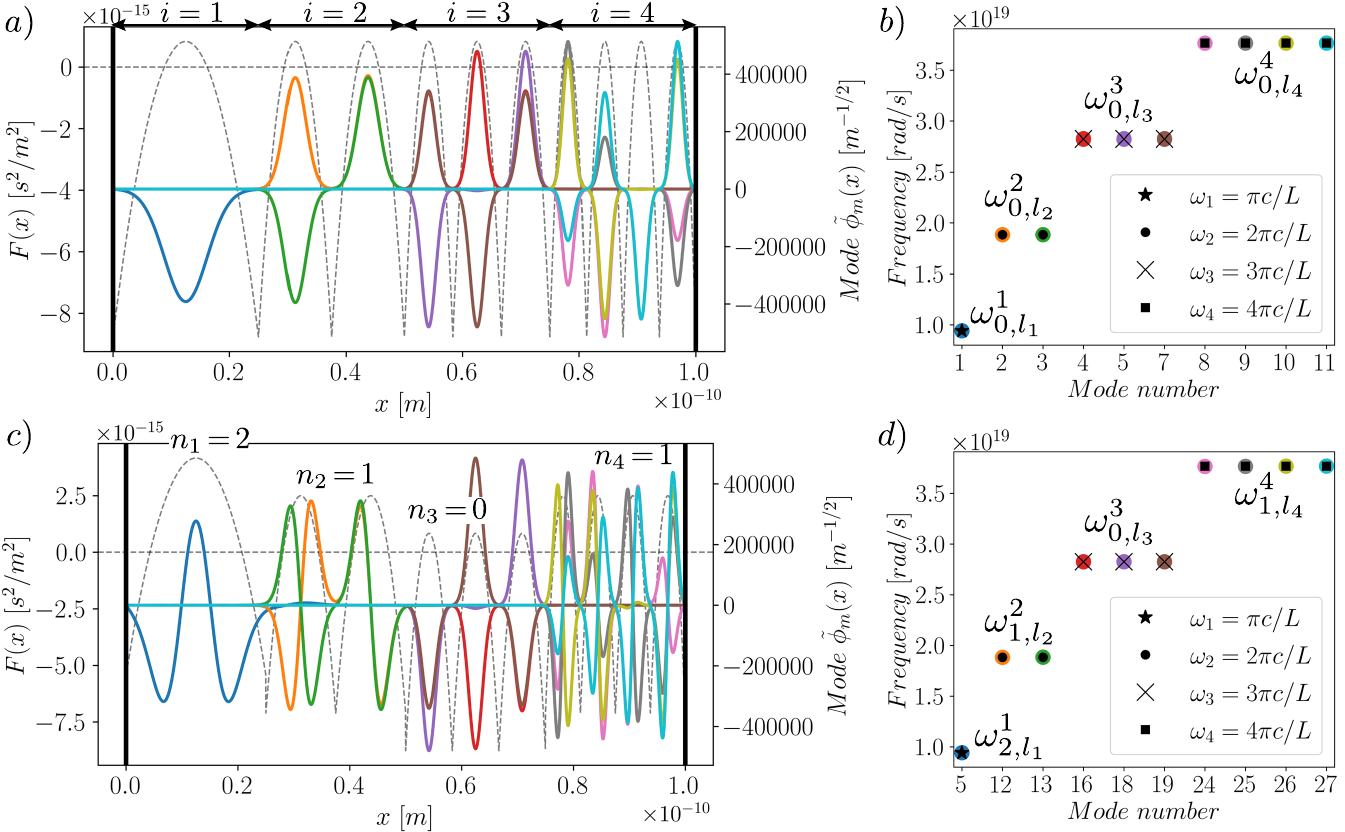}
	\caption{Restricting the Helmholtz coefficient $F(x)$ to be almost everywhere negative and the standing modes to the discrete QHO-like rules given in Eq.(\ref{eq:quantumstate}), electromagnetic stationary waves in a box can exhibit the spectrum of photons. a) Form of $F(x)$ with its four sub-cells $F_i(x)$ and the $N=10$ normalized modes $\tilde{\phi}_m(x)$ retained in the computed modal basis of Eq.(\ref{eq:waveelecbis}) to mimic the ``quantum ground state''. The box is of length $L=1\times 10^{-10}$ m and the $F(x)$ is 
    made of $i=1,\ldots,4$ concatenated $F_i(x)$ with $i$ wavelengths $\lambda_i=L/(4i)$, and a $C_0^i$ and $C_1^i$ given in Eq.(\ref{eq:quantumstate}) with $C_0=-100/c^2$ and $C_1=174.8/c^2$. b) Frequency spectrum of the $10$ modes shown in a). The first four frequencies of the theoretical photon spectrum $\omega_i=i\pi c/L$ are indicated by black markers. c) $F(x)$ with its four sub-cells $F_i(x)$ and the $10$ normalized modes $\tilde{\phi}_m(x)$ retained to mimic a ``quantum excited state''. The $F_i(x)$ and the associated modes $\phi_m(x)$ are such that $n_i=0$ for the sub-cell $i=1$, $n_i=2$ for $i=2$ and $n_i=1$ for $i=2$ and $i=4$. d) Frequency spectrum of the $10$ modes shown in c), corresponding to the first four photon frequencies shown in b).}
\label{fig:HQOemulator_wavesinabox_modes}
\end{figure}

To establish an analogy with photon quantization, we discretize the admissible functions $F(x)$ and the initial conditions $\tilde{q}_m(0)$. We consider a medium of length $L$ composed of $M$ concatenated sub-media $F_i(x)$, with $i=1,\ldots,M$, each containing $i$ wavelengths $\lambda_i = L/(Mi)$. The amplitudes $C_0^i$ and $C_1^i$ of the $i$-th QHO-like function $F_i(x)$ (see Eq.~(\ref{eq:periodmedia})) are constrained as
\begin{equation}
\label{eq:quantumstate}
C_0^i = (2n_i+1)(C_0+C_1)-C_1, \qquad
C_1^i = C_1, \qquad
\frac{\sqrt{C_1}}{C_0+C_1} = \frac{c}{4\sqrt{2}},
\end{equation}
where $n_i \in \mathbb{N}_0$ labels the selected band. We further impose a constant initial condition $\tilde{q}_m(0)=\tilde{E}_0$ only on modes $m$ belonging to the selected band $n_i$, while all other bands carry zero energy. To match the quantum energy scale, we deliberately define the amplitude $\tilde{E}_0$ in terms of the reduced Planck constant $\hbar$ as
\begin{equation}
\label{eq:e0hbar}
\tilde{E}_0=\tilde{q}^i_{n_i,\ell_i}(0)=\left(\frac{\mu_0\pi c^2\hbar}{2AL\sqrt{2C_1}}\right)^{\frac{1}{2}},
\qquad
i=1,\ldots,M,\;\; \ell_i=1,\ldots,i,\;\; n_i \in \mathbb{N}_0.
\end{equation}

\begin{figure}[htbp]
\centering
	\includegraphics[width=1\columnwidth]{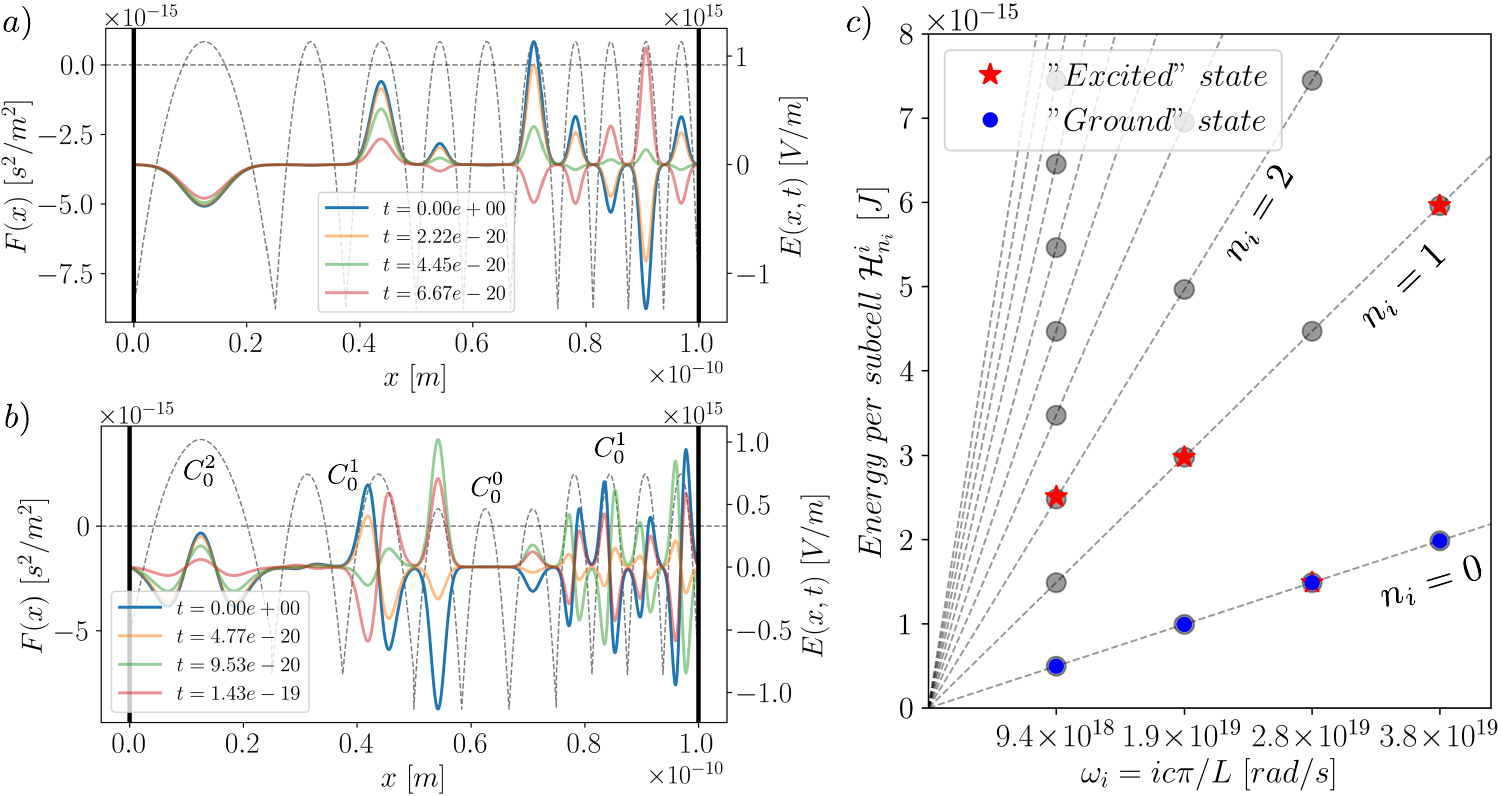}
	\caption{When subjected to an equipartition of energy, following Eq.(\ref{eq:e0hbar}), between the electromagnetic standing modes of Figure \ref{fig:HQOemulator_wavesinabox_modes}, the energy in the box can be described in a form analogous to photon energy.  a) Four time steps of the oscillatory electric field $E(x,t)=\sum_m\tilde{\phi}_m(x)\tilde{q}_m(t)$ with $\tilde{\phi}_m(x)$ and $F(x)$ shown in Figure \ref{fig:HQOemulator_wavesinabox_modes}a and $\tilde{q}_m(0)$ given in Eq.(\ref{eq:e0hbar}). b) As in a), but for the modal basis $\tilde{\phi}_m(x)$ and periodic medium $F(x)$ shown in Figure \ref{fig:HQOemulator_wavesinabox_modes}c). c) Energy $\mathcal{H}^i_{n_i}$ per sub-cell $F_i(x)$ as a function of frequency per sub-cell $\omega_i=i\pi c/L$. Blue disks and red stars show the energy of $E(x,t)$ in a) and b), respectively. The dotted lines and transparent black disks represent the quantum distribution of energy $\mathcal{E}=\sum_i (n_i+1/2)\hbar \omega_i$.}
\label{fig:HQOemulator_wavesinabox_energy}
\end{figure}

Under the assumptions Eq.~(\ref{eq:quantumstate}) and $\lambda_i=L/(Mi)$, the frequencies of the modes $\tilde{\phi}_m(x)=\tilde{\phi}^i_{n_i,\ell_i}(x)$ can be written as
\begin{equation}
\label{eq:omegaQHOter}
\omega_m=\omega^{i}_{n_i,\ell_i}=\left(n_i+\frac{1}{2}\right)\frac{2\pi\sqrt{2C^i_1}}{(C^i_0+C^i_1)\lambda_i}
=\frac{i\,4\pi\sqrt{2C_1}}{(C_0+C_1)L}
=\frac{i\pi c}{L}
\end{equation}
for $i=1,\ldots,M$, $\ell_i=1,\ldots,i$ \footnote{note that since $F_i(x)$ contains $i$ wavelengths $\lambda_i=L/(Mi)$, $P=i$ eigenfrequencies $\omega_m$ are obtained for each sub-medium}, and where $n_i \in \mathbb{N}_0$ labels the selected band. By construction, this expression reproduces the photon frequency spectrum; however, the discretization here originates from the sub-medium index $i$, rather than from global cavity modes as in the quantum case. As for the total energy in the box given in Eq.(\ref{Eq:classicalenerelecmodal}), following Eq.(\ref{Eq:energymodalt0}) it can be written as a sum over each sub-medium, $\mathcal{E}=\sum_{i=1}^Mi\mathcal{H}^i_{n_i}=A/(2\mu_0)\sum_{i=1}^Mi\tilde{m}^i_{n_i}\tilde{E}_0^2$ where $\tilde{m}^i_{n_i}$ is the effective mass given in Eq.(\ref{eq:finalHm}). Using Eqs.~(\ref{eq:quantumstate})-(\ref{eq:e0hbar}), the total energy becomes 
\begin{equation}
\label{eq:totalenerter}
\mathcal{E}=\sum_{i=1}^{M}i\mathcal{H}^i_{n_i}=\sum_{i=1}^{M}i\frac{A}{\mu_0}\tilde{E}_0^2\frac{C_0^i+C_1^i}{4}=\sum_{i=1}^{M}\left(n_i+\frac{1}{2}\right)i\frac{2A\tilde{E}_0^2\sqrt{2C_1}}{c\mu_0}=\sum_{i=1}^{M}\left(n_i+\frac{1}{2}\right)\hbar\omega_i
\end{equation}
with $\omega_i=i\pi c/L$. This expression is formally identical to the quantum energy spectrum of the electromagnetic field discussed in Section \ref{quantum2}, although here the correspondence is obtained within a classical wave framework through the medium-property and initial-condition assumptions introduced above.

We illustrate the results of Eqs.(\ref{eq:omegaQHOter})-(\ref{eq:totalenerter}) in Figures \ref{fig:HQOemulator_wavesinabox_modes} and \ref{fig:HQOemulator_wavesinabox_energy} for a box of length $L=1\times 10^{-10}$ m composed of $M=4$ concatenated sub-cells $F_i(x)$, with $C_0=-100/c^2$ and $C_1=174.8/c^2$ verifying Eq.(\ref{eq:quantumstate}). Figure \ref{fig:HQOemulator_wavesinabox_modes}a shows the four sub-functions $F_i(x)$ together with the ten normalized modes $\tilde{\phi}_m(x)=\tilde{\phi}^i_{0,\ell_i}(x)$, corresponding to $n_i=0$ for all $i$. The associated frequencies $\omega_m=\omega^i_{0,\ell_i}$, with $\ell_i=1,\ldots,i$ and $i=1,\ldots,M$, are shown in Figure \ref{fig:HQOemulator_wavesinabox_modes}b. They form the sequence $\omega_m = i\pi c/L$, each repeated $i$ times, in agreement with Eq.(\ref{eq:omegaQHOter}) and confirming the narrow-pass-band limit. Figures \ref{fig:HQOemulator_wavesinabox_modes}c-d show another configuration of $F_i(x)$ and the corresponding spectrum $(\tilde{\phi}^i_{n_i,\ell_i}(x),\omega^i_{n_i,\ell_i})$, where the selected band $n_i \in \mathbb{N}_0$ may take values $n_i>0$. As expected, the resulting frequency spectrum in Figure \ref{fig:HQOemulator_wavesinabox_modes}d is identical to that of Figure \ref{fig:HQOemulator_wavesinabox_modes}b and is analogous to the photon spectrum discussed in Section \ref{quantum2}.

Figure \ref{fig:HQOemulator_wavesinabox_energy}a shows four time steps of the evolution of the electric field $E(x,t)=\sum_m\tilde{\phi}_m(x)\tilde{q}_m(t)$, the solution of Eq.(\ref{eq:waveelecbis}). The initial conditions $\tilde{q}^i_{n_i,\ell_i}(0)$, given in Eq.(\ref{eq:e0hbar}), are imposed only on the modes $\tilde{\phi}^i_{0,\ell_i}(x)$ of band $n_i=0$ shown in Figure \ref{fig:HQOemulator_wavesinabox_modes}a.
As expected from Eq.(\ref{eq:totalenerter}), the total energy may be expressed as $\mathcal{E}=\sum_{i=1}^4i\mathcal{H}^i_{0}=\sum_{i=1}^4 \hbar\omega_i/2$, as shown by the blue disks in Figure \ref{fig:HQOemulator_wavesinabox_energy}c, which is analogous to a quantum ground state. When the parameters $(C_0^i,C_1^i)$ of Eq.(\ref{eq:quantumstate}) and the initial conditions are imposed on modes $\tilde{\phi}^i_{n_i,\ell_i}(x)$ with selected bands $n_i$ possibly greater than $0$, as illustrated in Figure \ref{fig:HQOemulator_wavesinabox_energy}b, the corresponding energy $\mathcal{E}=\sum_{i=1}^4i\mathcal{H}^i_{n_i}=(n_i+1/2)\hbar\omega_i$ is recovered, as shown by the red stars in Figure \ref{fig:HQOemulator_wavesinabox_energy}c. A slight deviation is observed for $\mathcal{H}^1_{2}$, since the localized mode $\tilde{\phi}_5(x)=\tilde{\phi}^1_{2,1}(x)$ is only weakly confined within $F_1(x)$, and the approximation $\int_0^L F_1(x)\tilde{\phi}_5(x)^2dx=(C_0^1+C_1^1)/2$ is not strictly satisfied. This highlights that the linear wave quantum analogue remains accurate only in the narrow-pass-band limit.

\section{Discussions and conclusions}
\label{conclu}

In this manuscript, we have theoretically investigated wave propagation in media that are periodically varying between an evanescent and propagative behavior. Using Floquet-Bloch theory and spectral analysis, we numerically characterized the formation of band gaps arising from Bloch-wave solutions of a one-dimensional wave equation with a Helmholtz coefficient $F(x)$ spatially varying between a negative and positive value. In this largely overlooked regime, the pass-band regions of the Bloch problem become sufficiently narrow that they may be regarded as forming an almost discrete set in parameter space. In this narrow-pass-band regime, the allowed propagating modes and their associated bands are governed by the spectrum of an effective Hamiltonian that is mathematically equivalent to that of a stationary Schr\"odinger equation. 

When Dirichlet boundary conditions are imposed, the resulting wave eigenvalue problem exhibits a discrete and nearly degenerate spectrum, where the multiplicity of eigenvalues and the associated phase-shifted eigenfunctions, as prescribed by the Schr\"odinger Hamiltonian, are determined by the number of wavelengths accommodated within the periodic medium. A remarkable consequence of this regime is the strong spatial localization of stationary wave modes within individual periods of the medium. As a result, the spectrum of a composite medium formed by concatenating sub-media is simply given by the union of the spectra associated with each constituent sub-medium. This property enables a powerful and modular design strategy, which we exploited to engineer a spatially varying photonic medium for electromagnetic waves in a one-dimensional cavity that exhibits an energy spectrum analogous to that predicted by the quantum theory of light.

Although our analysis is restricted to one-dimensional stationary waves in non-dispersive periodic media, it naturally extends to propagating waves, higher-dimensional geometries, and dispersive systems, all of which offer promising directions for future research. Owing to the universality of the wave equation with spatially varying Helmholtz coefficients, the present results are expected to apply broadly to classical linear waves, including mechanical, electrical, and photonic systems. 
Recent advances in fabrication have already enabled the realization of exotic passive metamaterials~\cite{schurig2006electric,huang2009negative,cummer2016controlling,yves2017crystalline,fan2022active} and active structural media~\cite{fleury2014negative,wang2017active,veenstra2024non} exhibiting effective negative properties, which provide a natural platform for realizing the narrow pass-bands described in this work. Another promising route lies in slender elastic structures subjected to spatially varying compressive stresses near buckling~\cite{read2025wave}, which can exhibit narrow wavenumber lock-in regions in modulation space closely related to the emergence of the band gaps considered here in the context of wave dynamics.

We anticipate that the pass-band thinning mechanism presented here can help narrow the conceptual gap between classical and quantum wave theories by showing how quantization-like spectra may emerge within a purely linear framework. In this sense, it complements a growing body of attempts to rationalize and analogize quantum phenomena classically across a wide range of systems, including hydrodynamic~\cite{couder2005dynamical,bush2015pilot,rozenman2019amplitude,bush2021hydrodynamic,Deymier2026, JoshuaA2026}, acoustic~\cite{hasan2019sound,Deymier2026} and mechanical~\cite{neder2024bloch,lazarus2025optimal,JoshuaA2026} quantum analogs, stochastic electrodynamics~\cite{de2013quantum,boyer2019stochastic} and light-matter interactions~\cite{weisman2021diffractive,goncalves2022bright}.

It may be tempting to interpret the Maxwell equation with the periodic Helmholtz coefficient $F(x)$ introduced in this manuscript as providing direct physical insight into photon quantization. However, such an interpretation would be premature, as $F(x)$ is imposed \emph{a priori} and does not arise from genuine light-matter interactions. A natural and compelling direction for future work is therefore to investigate whether Floquet-type stabilization and pass-band narrowing can emerge from particle-field coupling in the Maxwell-Lorentz equations, for instance through dynamically stabilized lattices of unstable particles in external or self-consistent fields. If such mechanisms were shown to exist, they could offer a concrete pathway toward bridging classical light-matter interactions and the emergence of quantum-mechanical phenomena.

\enlargethispage{20pt}


\section*{Author Contributions}
A.L. conceived and coordinated the study, developed the physical framework, performed the quantum derivations, conducted the spectral analysis of waves propagating in periodic media, developed the mathematics of the Floquet--Bloch quantum analogues, carried out the numerical simulations, and wrote the manuscript. G.G.R. contributed through editing and review of the quantum derivations. J.W.M.B. coordinated the study, acquired funding, administered the project, contributed to writing, and reviewed and edited the manuscript.
\section*{Competing Interests}
The authors declare no competing interests.
\section*{Funding}
A.L. and J.W.M.B. acknowledge support from the Office of Naval Research under grant N000014-24-1-2232. J.W.M.B. acknowledges support from the National Science Foundation under grant CMMI-2154151. G.G.R. acknowledges additional support from the MIT School of Science Research Innovation Seed Fund and the C.L.E. Moore Instructorship.
\section*{Data availability}
The numerical code used to produce all figures in this article is freely available on GitHub at \url{https://github.com/nauradsurazala/wave-energy-quantization} and has been archived on Zenodo at \url{https://doi.org/10.5281/zenodo.20706188}.





\vskip2pc




\clearpage
\setcounter{section}{0}
\setcounter{figure}{0}
\numberwithin{equation}{section}
\renewcommand{\thefigure}{S\arabic{figure}}
\renewcommand{\thetable}{S\arabic{table}}
\titleformat{\section}{\normalfont\bfseries}{S\thesection.}{0.5em}{}
\titleformat{\subsection}{\normalfont\itshape}{S\thesubsection.}{0.5em}{}
\raggedbottom
\needspace{8\baselineskip}
\begin{center}
  {\normalsize\textbf{Electronic Supplementary Material}}\\[0.6em]
  {\large Engineering classical waves with quantized energy spectra in periodic media}\\[0.9em]
  Arnaud Lazarus$^{1,2}$, Georgi Gary Rozenman$^{1}$ and John W. M. Bush$^{1}$\\[0.5em]
  {\small $^{1}$ Department of Mathematics, Massachusetts Institute of Technology, Cambridge, MA, USA\\
  $^{2}$Institut Jean Le Rond d'Alembert, CNRS UMR7190, Sorbonne Universit\'e Paris, France}
\end{center}

\vspace{1em}

\section{Canonical quantization of light energy in a $1D$ empty cavity}
\label{Supp:quantum2}

We have shown in Section 2 of the manuscript that  the Hamiltonian of stationary modes of light in an empty cavity takes a well-known form analogous to that of a harmonic oscillator \cite{dragoman2004quantum}
\begin{equation}
\mathcal{H}_m = \frac{1}{2}k_mq_m^2(t) + \frac{p_m^2}{2m_m}
\label{Supp:Eq:totalclassicalfinal}
\end{equation}
when expressed in its classical orthogonal modal basis.

To mathematically obtain the quantization of Eq.(\ref{Supp:Eq:totalclassicalfinal}), one needs to quit the classical world to enter the quantum one, which provides a natural formalism for obtaining quantization \cite{schleich2015quantum}. The process starts by promoting the canonical variables $q_m$ and $p_m$ to operators $\hat{q}_m$ and $\hat{p}_m$ with the canonical commutation relation
\begin{equation}
\left[\hat{q}_m,\hat{p}_m\right]=\hat{q}_m\hat{p}_m-\hat{p}_m\hat{q}_m=i\hbar \delta_{mn}
\label{Eq:canonicalcommutation}
\end{equation}
We then introduce the ladder operators, the so-called annihilation $\hat{a}_m$ and creation $\hat{a}^{\dagger}_m$ operators 
\begin{equation}
\left\{
\begin{split}
\hat{a}_m = \sqrt{ \frac{m_m\omega_m}{2 \hbar} } \left( \hat{q}_m + \frac{i}{m_m\omega_m}\hat{p}_m \right), \\
\hat{a}^\dagger_m = \sqrt{ \frac{m_m \omega_m }{2\hbar} } \left(\hat{q}_m-\frac{i}{m_m\omega_m}\hat{p}_m \right).\\
\end{split}
\right.
\label{Eq:ladderoperator}
\end{equation}
that verify the commutation relation $\left[\hat{a}_n,\hat{a}^\dagger_m\right]=\delta_{nm}$. From Eq.(\ref{Eq:ladderoperator}), it is possible to express the canonical operators in terms of ladder operators, following
\begin{equation}
\left\{
\begin{split}
\hat{q}_m = \sqrt{ \frac{\hbar}{2 m_m\omega_m} } \left( \hat{a}_m + \hat{a}_m^\dagger \right), \\
\hat{p}_m = -i \sqrt{ \frac{ \hbar m_m \omega_m }{2} } \left(\hat{a}_m-\hat{a}_m^\dagger \right).\\
\end{split}
\right.
\label{Eq:canonicaloperator}
\end{equation}
The operator form of the Hamiltonian $\mathcal{H}_m$ given in Eq.(\ref{Supp:Eq:totalclassicalfinal}) then becomes
\begin{equation}
\hat{\mathcal{H}}_m = \frac{1}{2}k_m\hat{q}_m^2(t) + \frac{\hat{p}_m^2}{2m_m}=\hbar\omega_m\left(\hat{a}_m^\dagger\hat{a}_m + \frac{1}{2}\right)
\label{Eq:Hamiltonianoperator}
\end{equation}
that is the standard Hamiltonian operator of the Quantum Harmonic Oscillator.

We work in the Fock basis, which is the natural basis when one is interested in photon counting, as is the case in this manuscript. The fundamental elements of the Fock basis are the number states $|n_m\rangle$, which represent eigenstates of the Hamiltonian $\hat{\mathcal H}_m$ associated with mode $m$, containing an integer number $n_m \in \mathbb{N}_0$ of photons. These states form an orthonormal basis of the Hilbert space of the quantized field. The dual vectors $\langle n_m|$ are the Hermitian conjugates of $|n_m\rangle$, and satisfy $\langle n_m | n_m' \rangle = \delta_{n_m n_m'}$.
By definition, the number operator for mode $m$,
\begin{equation}
\hat{N}_m := \hat{a}_m^\dagger \hat{a}_m,
\end{equation}
acts on the Fock states as
\begin{equation}
\hat{N}_m |n_m\rangle = n_m |n_m\rangle.
\end{equation}
The Hamiltonian operator associated with mode \(m\) is given by
\begin{equation}
\hat{\mathcal H}_m = \hbar \omega_m \left( \hat{N}_m + \tfrac{1}{2} \right),
\end{equation}
and is diagonal in the Fock basis. Acting on a number state \(|n_m\rangle\), it yields
\begin{equation}
\hat{\mathcal H}_m |n_m\rangle
= \hbar \omega_m \left( n_m + \tfrac{1}{2} \right) |n_m\rangle .
\label{Eq:Hamiltonianactsonn}
\end{equation}
The corresponding expectation value of the energy of mode \(m\) in the state \(|n_m\rangle\) is therefore
\begin{equation}
\mathcal H_m^{(n_m)}
= \langle n_m | \hat{\mathcal H}_m | n_m \rangle
= \hbar \omega_m \left( n_m + \tfrac{1}{2} \right).
\label{Eq:Energyonemode}
\end{equation}
Finally, the total energy of the quantized electromagnetic field, obtained by summing over all independent modes, reads
\begin{equation}
\mathcal E
= \sum_m \mathcal H_m
= \sum_{m=1}^{\infty} \hbar \omega_m \left( n_m + \tfrac{1}{2} \right),
\label{Supp:Eq:totalclassicalfinalquantum}
\end{equation}
which is the standard expression for the energy of a quantized field in terms of its mode occupations.


\newpage
\section{Classic propagation of a $1D$ wave in an infinite periodic media}
\label{inflong}

In this section, we study the classic case of a wave propagating in a one-dimensional, periodic, continuous, medium that is modeled by the generic Helmholtz equation~\cite{brillouin1946wave,richards2012analysis}: 
\begin{equation}
\label{Supp:Helmholtz}
\frac{\partial^2 u(x,t)}{\partial x^2} + \omega^2 F(x)u(x,t) = 0
\end{equation}
where $x$ and $t$, represent space and time, respectively, $u(x,t)=\phi(x)e^{-i\omega t}$ is the stationary wave oscillating at frequency $\omega$ with spatial shape $\phi(x)$ and $F(x)$ is a $\lambda$-periodic function where $\lambda$ is the wavelength of the structured medium. In this Section, we consider an infinitely long periodic medium. The influence of boundary conditions will be treated in Section S\ref{notinflong}. Probably the most iconic case of Eq.(\ref{Supp:Helmholtz}) is when the modulation is harmonically varying in the form
 \begin{equation}
\label{Supp:Harmonicmod}
F(x)=C_0+C_1\cos(2\pi x/\lambda)
\end{equation} 
with $C_0$ and $C_1$ two real numbers (whose units are homogeneous to one over velocity squared) representing the natural and modulated part of the medium property. In the case where $C_0=1/c^2$, with $c$ for example the speed of light for electromagnetic waves in vacuum, and $C_1 \rightarrow 0$, Eq.(\ref{Supp:Helmholtz}) is simply the Helmoltz equation of a classic wave equation giving a dispersion relation $\omega=c k$ where $k$ is the wavenumber of the wave.


\begin{figure}[!htb]
\centering
	\includegraphics[width=1\columnwidth]{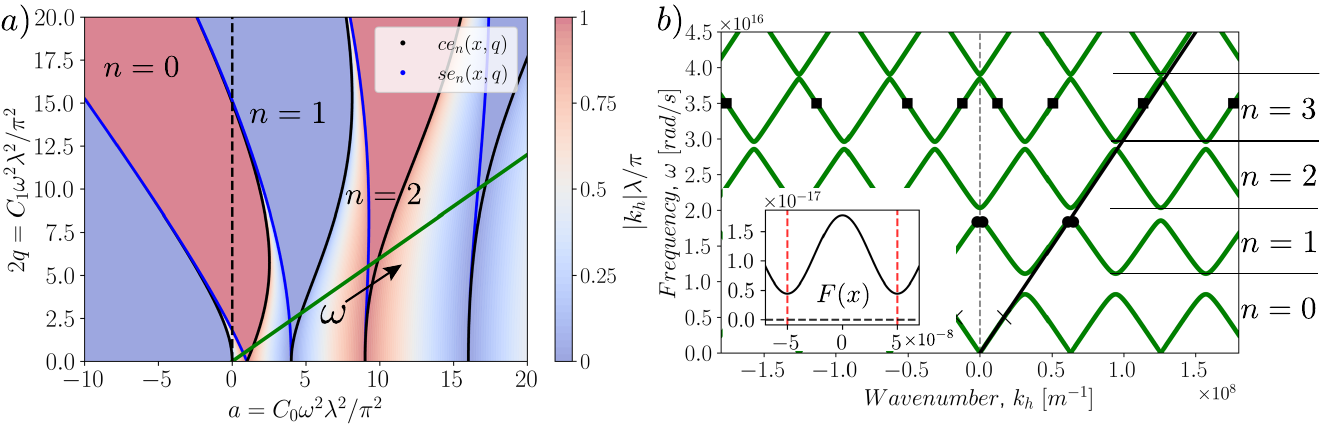}
	\caption{Computation of the wave dispersion curve of a $1D$ wave in a periodically modulated continuous medium. a) Evolution of the Bloch wave dimensionless wavenumber $|k_h|\lambda/\pi$ belonging to the first Brillouin zone $-\pi/\lambda \leq k_h \leq \pi/\lambda$ in the $(a,2q)$ modulation space. The case $\eta=C_1/C_0=0.6$, illustrated in the inset of b), is represented by a green line. The  tongues associated to non-propagating Bloch waves with a non-zero Floquet exponent are shown with plain colors and denoted by the integer $n$. f) Extended wave dispersion relation of a wave in a periodic media with $F(x)$ given in inset with $C_0=1/c^2$ and $\lambda=1 \times 10^{-7}$ m The classic dispersion relation $\omega=k/\sqrt{C_0}=ck$ for a medium with no modulation, i.e. $C_1 \rightarrow 0$, is indicated by the black line. We highlight the periodicity of the frequency in the wavenumber $k_h = k + h\pi/\lambda$ with $h \in \mathbb{Z}$ by plotting vertical dashed lines every $\pi/\lambda$. The location of the dominant peaks of the wavenumber spectrum of the three allowed waves shown in Figure~\ref{Supp:Figure2_stationary_modes_classic} is displayed with crosses, disks and squares for a frequency in the passing band $n=0$, $1$ and $3$, respectively.}
\label{Supp:Figure1_classicdispersionrelation}
\end{figure}

Introducing the dimensionless variable $\xi=\pi x/\lambda$ and dropping the oscillating part in $u(x,t)=\phi(x)e^{-i\omega t}$,  Eq.~(\ref{Supp:Helmholtz}) can be recast in the shape of the well-known Mathieu equation
\begin{equation}
\label{Supp:Mathieu}
\frac{d^2 \phi(\xi)}{d \xi^2} + \left[a + 2q\cos(2\xi))\right]\phi(\xi) = 0
\end{equation}
with $a=\omega^2\lambda^2C_0/\pi^2$ and $2q=\omega^2\lambda^2C_1/\pi^2$ so that $a$ and $q$ are related by $a=2 q/\eta$ where $\eta=C_1/C_0$ represents the relative importance of modulation. We recall that according to Floquet theory, we know that the solution $\phi(\xi)$ takes the  form
\begin{equation}
\label{Supp:Floquetform}
\phi(\xi) = A\Psi(\xi)e^{s\xi} + B\overline{\Psi}(\xi)e^{-s\xi}\end{equation}
where $\Psi(\xi)$ is a complex $\pi$-periodic function with the same period as $F(\xi)$, $\overline{\Psi}(\xi)$ is its complex conjugate, $s$ is a complex number called the Floquet exponent, and $A$ and $B$ are arbitrary constants. 

\begin{figure}[!b]
\centering
	\includegraphics[width=1\columnwidth]{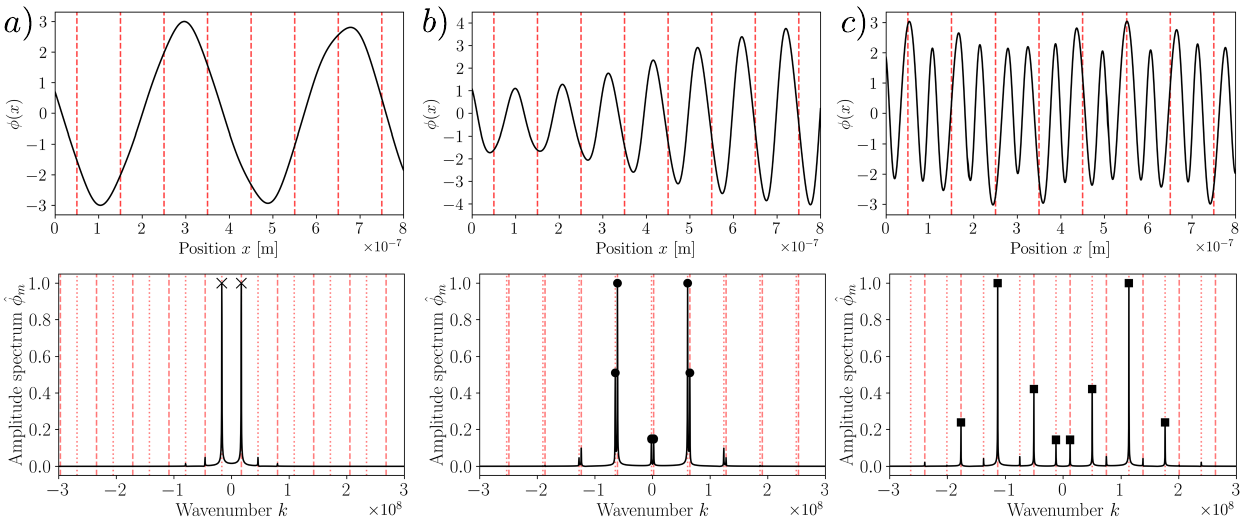}
	\caption{Spatial part $\phi(x)$ of the wave $u(x,t)=\phi(x)e^{-i\omega t}$ (top row) and their wavenumber spectrum (bottom row) for $F(x)=C_0+C_1\cos(2\pi x/\lambda)$ with $\lambda=1\times 10^{-7}$ m, $C_0=1/c^2$ and $\eta=C_1/C_0=0.6$ as shown in the inset of Figure~\ref{Supp:Figure1_classicdispersionrelation}b. Dashed red vertical lines in the spatial domain are separated by a wavelength $\lambda$ and locate the minima of the cosine function $F(x)$. In the reciprocal wavenumber domain, the vertical dashed red lines indicate the wavenumbers location of $\phi(x)$'s spectrum, that reads $k_h=k+h2\pi/\lambda$ with $h \in \mathbb{Z}$. a) Example of a wave with a frequency $\omega$ in the first pass band $n=0$ as shown by black crosses in Figure~\ref{Supp:Figure1_classicdispersionrelation}b. b) Wave with a frequency $\omega$ in the second pass band $n=1$ as shown by black disks in Figure~\ref{Supp:Figure1_classicdispersionrelation}b. c) Wave with a frequency $\omega$ in the fourth pass band $n=3$ as shown by black squares in Figure~\ref{Supp:Figure1_classicdispersionrelation}b.}
\label{Supp:Figure2_stationary_modes_classic}
\end{figure}

To gain physical insights in the wave problem, one computes the dispersion relation that is the relation between $\omega$ and the wavenumber spectrum of $\phi(\xi)$ for a given $F(x)$, i.e. for a fixed $\eta=C_1/C_0=2q/a$. We recall that since $\Psi(\xi)$ is $\pi$-periodic, it can be expanded in a Fourier series $\Psi(\xi)=\sum_{h=-\infty}^{\infty}\Psi_he^{i2h}$ so that the actual wavenumbers contained in the spectrum of the Floquet form $\phi(\xi)$ read $s_h=\sigma +i\kappa_h$ where $\kappa_h = \kappa + 2h$ with $h \in \mathbb{Z}$ is the reciprocal lattice in Bloch theorem. Figure~\ref{Supp:Figure1_classicdispersionrelation}a, already shown in the manuscript, is well-known in the framework of waves propagating in periodic media and shows the evolution of the wavenumber $|\kappa_h|$ belonging to the so-called first Brillouin zone $-1 \leq \kappa_h \leq 1$, as a function of $a$ and $2q$. As already explain in the manuscript, the plain colored regions are stop bands corresponding to the parameters $(a,q)$ for which $\sigma \neq 0$ and attenuation occurs (blue and red regions correspond to $s_h=\sigma + i2h$ and $s_h=\sigma + i(1+2h)$, respectively). The multicolored regions are passing bands for which $\sigma=0$ and $s$ are purely imaginary. Mathematically, the curves $a=f(q)$ separating between graded and plain colored regions are defined by the so-called Mathieu characteristic values. Physically, those pass-band limits correspond to modulation parameters $(a,q)$ for which a Bragg resonance condition occurs.

In this section, we assume $\eta=C_1/C_0=2q/a=0.6$ which is a typical case of wave propagating in periodic media where broad pass or transmission bands and narrow stop or absorption bands occur. In this regime corresponding to $F(x)>0$ $\forall x$, $q \rightarrow 0$ corresponds to the limit of a continuous medium with no periodicity. Since $a=\omega^2\lambda^2C_0/\pi^2$ and $2q=\omega^2\lambda^2C_1/\pi^2$, increasing the frequency $\omega$ for a given medium is represented by a straight line with slope $\eta$ running from the origin of the $(a,2q)$ space as illustrated by the green line of figure~\ref{Supp:Figure1_classicdispersionrelation}a. As $\omega$ increases, this line cuts successive large pass bands of $\sigma = 0$ whose order is identified by the integer $n=0,1,\ldots$. The resulting extended dispersion curve is given in figure~\ref{Supp:Figure1_classicdispersionrelation}b, showing the propagating frequencies $\omega$, associated with $\sigma = 0$, as a function of $k_h=\kappa_h\pi/\lambda$ for $F(x)=C_0+ C_1\cos(2\pi x/\lambda)$ shown in the inset of Figure~\ref{Supp:Figure1_classicdispersionrelation}b with $\lambda=1 \times 10^{-7}$ m and  $C_0=1/c^2=C_1/0.6$ where $c$ is chosen to be the speed of light. The black line $\omega=k/\sqrt{C_0}$ that is, in our particular case, $\omega=ck$, has been plotted for comparison with a classic dispersion curve in a homogeneous medium with $F(x)=1/c^2$. The wavenumber $k$ of the extended dispersion relation $\omega = f(k_h)$ that is the closest to the line $\omega=ck$ indicates the location of the fundamental peak in the wavenumber spectrum of $\phi(x)$ as shown in the bottom row of Figure~\ref{Supp:Figure2_stationary_modes_classic}.


The top row of Figure~\ref{Supp:Figure2_stationary_modes_classic} shows the spatial evolution of $\phi(x) = A\Psi(x)e^{ikx} + B\bar{\Psi}(x)e^{-ikx}$ with $\Psi(x)=\Psi(x+\lambda)$ and with an associated frequency $\omega=0.50\times 10^{16}$ rad/s, $\omega=1.84\times 10^{16}$ rad/s and $\omega=3.5\times 10^{16}$ rad/s located in the first, second and fourth passing band regions $n$ of Figure~\ref{Supp:Figure1_classicdispersionrelation}b, respectively. The bottom row of the figure shows the corresponding wavenumber spectrum that has been computed via a Fast Fourier Transform (FFT) of $\phi(x)$, following
\begin{equation}
\label{eq:FFT}
\hat{\phi}(k)=\frac{1}{N}\sum^{N-1}_{j=0}\phi(x_j)e^{-i2\pi kj/N}\end{equation}
where the total number of sample points is $N=100000$ over $100$ wavelengths (the FFTs have been normalized so that the maximum amplitude is $max(\hat{\phi}(k))=1$). 

As expected from the Floquet  form given in Eq.(\ref{Supp:Floquetform}), the location of the wavenumber peaks coincides with the imaginary part of the Floquet exponents $k_h=\pm k+h2\pi/\lambda$ that are plotted in red vertical dashed lines on the wavenumber spectra. The location of the predominant peaks with $\hat{\phi}(k)>0.1$ illustrated by crosses, disks and square markers for $\omega_0$, $\omega_1$ and $\omega_3$, respectively, has been reported on the dispersion curve of Figure \ref{Supp:Figure1_classicdispersionrelation}b to highlight the connection between the frequency of the waves and their wavenumber spectrum. Because we consider the infinitely long problem with a Bloch-wave approximation of the solutions, as $a$ and $q$ continuously span the passing bands, an infinite number of solutions $u(x,t)$ exist that are associated with Floquet multipliers, $\rho = e^{ik\lambda}$ and $\bar{\rho} = e^{-ik\lambda}$ continuously varying on the  imaginary unit circle such that $|\rho|=|\bar{\rho}|=1$. The stationary waves of Figure~\ref{Supp:Figure2_stationary_modes_classic} were randomly selected in their passing band $n$ such that we had $\rho=-0.111+0.994i$, $\rho=0.980+0.200i$ and $\rho=0.346+0.938i$ for Figures~\ref{Supp:Figure2_stationary_modes_classic}a, \ref{Supp:Figure2_stationary_modes_classic}b and \ref{Supp:Figure2_stationary_modes_classic}c, respectively. Note that $\omega$ in the $n=1$ pass-band region has been chosen for an $(a,q)$ close to the Bragg resonance conditions, i.e. close to the limit of the pass band, where a beating phenomenon occurs (see Figure~\ref{Supp:Figure2_stationary_modes_classic}b). This beating phenomenon arises because the two Floquet multipliers of $\phi(x)$ are almost locked-in to $\rho=\bar{\rho}=1$, which correspond to a purely, possibly amplified, periodic mode, hence the term Bragg  ``resonance'' at the very end of each pass band.

\newpage
\section{FFT of bloch waves in the narrow-pass-band regime}

\begin{figure}[!h]
\centering
	\includegraphics[width=1\columnwidth]{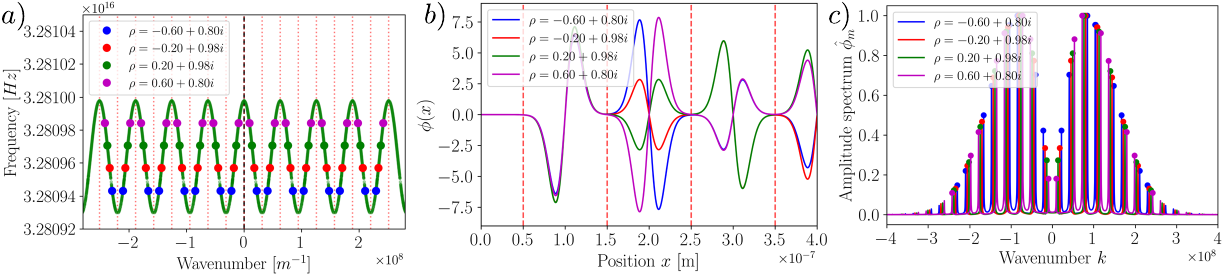}
	\caption{Example of four waves whose frequencies are equally spaced in the narrow pass band $n=1$ for $\lambda=1 \times 10^{-7}$ m and $\eta=-3$. a) Zoom on the dispersion relation of Figure 2f of the manuscript for $n=1$. The equally spaced frequencies correspond to equally spaced Floquet multipliers $\rho$ on the unit imaginary circles, i.e. $|\rho|=1$. b) Evolution of the spatial part of the waves $\phi(x)$ as a function of distance $x$. The dashed vertical lines locate the minima of the cosine function $F(x)$ shown in Figure~2a of the manuscript. c) FFTs $\bar{\phi}(k)$, normalized so that $max(\bar{\phi}(k))=1$, of the four waves $\phi(x)$. Color dots indicate the amplitudes of the predominant peaks such that  $\bar{\phi}(k)>0.1$. Their location is plotted in the corresponding dispersion curve in a).}
\label{fig:multimodesTransmon}
\end{figure}

We recall that the propagating stationary waves shown in Figures~1b-d of the manuscript are not the only ones in the narrow pass bands of Figure 1f. Actually, when we consider the infinitely long problem, as $a$ and $q$ continuously span the pass bands, an infinite number of Bloch waves $u(x,t)$ exist that are associated with Floquet multipliers, $\rho = e^{ik\lambda}$ and $\bar{\rho} = e^{-ik\lambda}$ continuously varying on the imaginary unit circle such that $|\rho|=|\bar{\rho}|=1$. The stationary waves of Figures~1b-d of the manuscript were randomly selected in their pass band $n$ such that $|\rho|=1$. Figure~\ref{fig:multimodesTransmon}b shows four stationary Bloch waves whose frequencies are equally spaced in the pass band $n=1$ that has been zoomed in and  illustrated in Figure~\ref{fig:multimodesTransmon}a. The  Floquet multipliers $\rho$ of those Bloch waves are equally spaced on the imaginary unit circle and their wavenumber spectrum, illustrated by the FFTs of $\phi(x)$ in Figure~\ref{fig:multimodesTransmon}c are simply shifted by $k_h=\pi/\lambda(\kappa+2h)$. 

Although each stationary wave $\phi(x) = A\Psi(x)e^{ikx} + B\Psi(x)e^{-ikx}$ associated with various $\rho$ such that $|\rho|=1$ looks different in the physical space, as shown in Figure~\ref{fig:multimodesTransmon}b, 
each $\phi(x)$ is actually a succession of a scaled copy of a single $\Psi(x)=\Psi(x+\lambda)$ over a modulation wavelength $\lambda$. Also in the reciprocal space like in Figure~\ref{fig:multimodesTransmon}b, the global shape of the FFT and the spectral energy, i.e. the distribution of harmonic amplitudes, is conserved and depends on the narrow pass-band we are considering. These properties are not common in Floquet systems, but start to emerge in the narrow-pass-band limit when the functions $\phi(x)$ solutions of Eq.(\ref{Supp:Mathieu}) become compact on each wavelength as already mentioned in~\cite{lazarus2019discrete,grandi2023new,lazarus2025meaningful,lazarus2025optimal} in the framework of dynamical systems. 


\newpage
\section{QHO limit in the Transmon}

\begin{figure}[!b]
\centering
	\includegraphics[width=1\columnwidth]{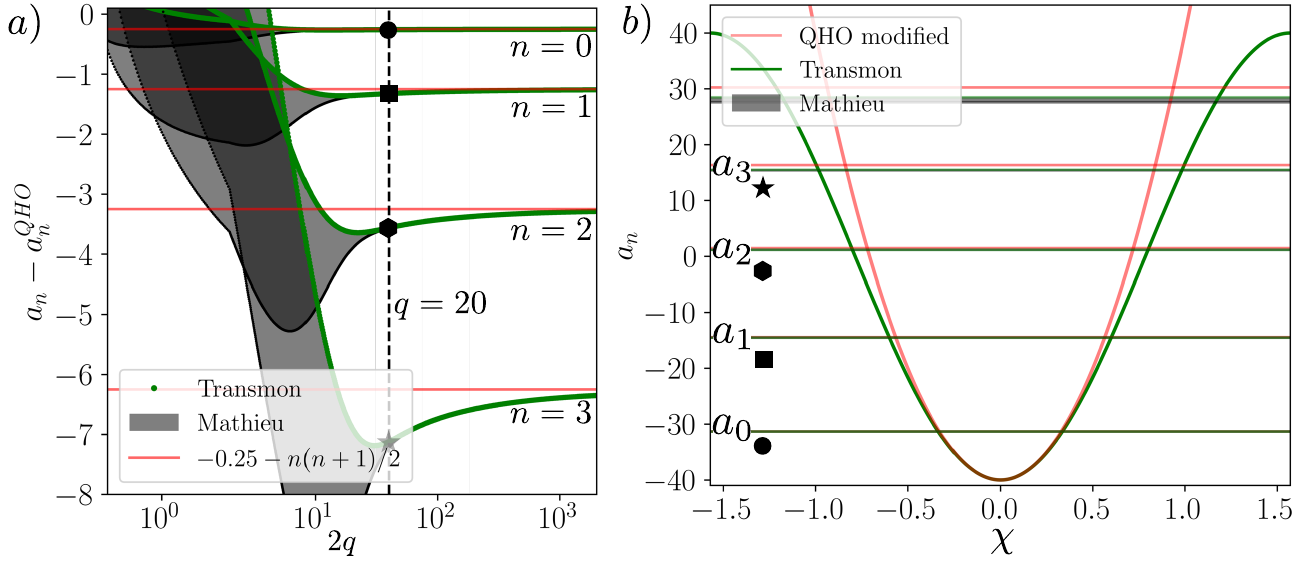}
	\caption{Approximation of the Mathieu equation by a stationary Schr\"odinger equation in the narrow-pass-band limit. a) First four pass-band regions $a_n-a_n^{QHO}$ as a function of $2q$ where $a_n^{QHO}=(n+1/2)4\sqrt{q}-2q$ is the QHO approximation. The eigenvalues $a_n$ of Eq.(\ref{Supp:Transmon}) are shown with green dots. The horizontal red lines are $a_n-a_n^{QHO}=-\frac{n(n+1)}{2}-0.25$. b) Green horizontal lines show the typical representation of the first four eigenvalues $a_n$ of Eq.(\ref{Supp:Transmon}) with a dimensionless potential $\bar{\mathcal{E}}=-2q\cos(2\xi)$ with $q=20$. The stability regions $a_n$ of the Mathieu equation Eq.(\ref{Supp:Mathieu}) for $q=20$ are reported with filled gray regions. The  red lines are the $a_n^{QHO}=(n+1/2)4\sqrt{q}-2q$, eigenvalues of Eq.(\ref{Supp:QHO}) for $q=20$.}
\label{Supp:fig:reductiontransmon}
\end{figure}

Let's denote $\Psi_n(\xi)=\phi^{\rho=1}_n(\xi)$, with $n=0,1,2,\ldots$, the $\pi$-periodic solutions of Eq.(\ref{Supp:Mathieu}) with $\rho=1$ which are located in the $(a,q)$ space at the frontier of the $n^{th}$ order narrow pass band as shown in Figure~\ref{Supp:Figure1_classicdispersionrelation}a by the lines of $|\kappa_h|=0$ surrounding the blue Mathieu tongues. Because Eq.(\ref{Supp:Mathieu}) is a Mathieu equation, $\phi^{\rho=1}_n(\xi)$ are simply even Mathieu functions $ce_{n}(\xi,q)$ for even integers $n$ and odd Mathieu functions $se_{n+1}(\xi,q)$ for odd integers $n$. Consequently, the  discrete set of parameters $a_n(q)$ leading to $\rho=1$ are the associated Mathieu characteristic values $a_n(q)$ and $b_n(q)$.

For a given dimensionless modulation parameter $q$, the solutions $\Psi_n(\xi)=\phi^{\rho=1}_n(\xi)$ and their associated $a_n$ are not only solutions of the Mathieu Eq.(\ref{Supp:Mathieu}) but of the eigenvalue problem
\begin{equation}
\label{Supp:Transmon}
\left[-\frac{d^2}{d\chi^2} -2q\cos(2\chi)\right]\Psi_n(\chi)=a_n\Psi_n(\chi),
\end{equation}
where $\chi \in [-\pi/2,\pi/2]$. Eq.(\ref{Supp:Transmon}) is simply a rewrite and change of variable of Mathieu equation. 
For a given $q$, the eigenvalues of Eq.(\ref{Supp:Transmon}) are the  characteristic values $a_n$ associated with $\pi$-periodic Mathieu functions. In this form over $[-\pi/2,\pi/2]$, Eq.(\ref{Supp:Transmon}) is well-known in the physics community because it is the dimensionless form of the stationary Schr\"odinger equation of a transmon Qubit \cite{roth2022transmon}.

Figure \ref{Supp:fig:reductiontransmon}a shows in green the evolution of the first four eigenvalues $a_0$, $a_1$, $a_2$ and $a_3$ of Eq.(\ref{Supp:Transmon}), as a function of $2q$. To compare with a well-known analytical formula from quantum mechanics, we plot the difference $a_n-a_n^{QHO}$ where $a_n^{QHO}+2q$ are the eigenvalues of 
\begin{equation}
\label{Supp:QHO}
\left[-\frac{d^2}{d\chi^2}+4q\chi^2\right]\Psi_n(\chi)=(a_n^{QHO}+2q)\Psi_n(\chi)
\end{equation}
that has been obtained by approximating $2q\cos(2\chi)$ in Eq.(\ref{Supp:Transmon}) by its Taylor expansion $2q\cos(2\chi) = 2q(1-2\chi^2)$. The QHO limit given in Eq.(\ref{Supp:QHO}) is mathematically equivalent to the stationary part of the Schr\"odinger equation of a Quantum Harmonic Oscillator (QHO) with quadratic potential $4q\chi^2$. The expression of QHO eigenvalues is known in the closed form~\cite{messiah1961quantum} 
\begin{equation}
\label{Supp:eigenQHO}
a_n^{QHO}+2q=(n+1/2)4\sqrt{q}.
\end{equation} 
As $q$ increases and the approximation $\cos(2\chi)\Psi_n(\chi) = (1-2\chi^2)\Psi_n(\chi)$ becomes more accurate due to the increasing localization of $\Psi_n(\chi)$ around $\chi=0$, the eigenvalues $a_n$ of the Transmon equation Eq.(\ref{Supp:Transmon}) converge towards those of the QHO. However, a finite correction (one speaks of a remaining anharmonicity in the QHO limit of the Transmon), so we eventually get $a_n-a_n^{HQO}=-\frac{n(n+1)}{2}-0.25$ as illustrated by the red horizontal lines in Figure \ref{Supp:fig:reductiontransmon}a. 

An interesting observation can be made by plotting the first four pass-band regions of Figure \ref{Supp:Figure1_classicdispersionrelation}a in Figure \ref{Supp:fig:reductiontransmon}a. As expected, the characteristic values $a_n$ of Eq.(\ref{Supp:Transmon}) associated with the $\pi$-periodic Mathieu functions, displayed by green dots in Figure \ref{Supp:fig:reductiontransmon}a, are the boundaries of the pass bands. An often overlooked feature is that as $q$ increases and the narrow-pass-band assumption becomes reasonable, the whole pass-band region shrinks towards the eigenvalues $a_n$ of the stationary Schr\"odinger equation Eq.(\ref{Supp:Transmon}). Figure \ref{Supp:fig:reductiontransmon}b, which is a typical representation of the energy levels of quantum bound states of a Transmon, shows that at $q=20$ for example, one gets already a good approximation of the location of the first four whole pass-band regions (horizontal gray regions) by computing the first eigenvalues of Eq.(\ref{Supp:Transmon}) illustrated by horizontal green lines. As a reference, the QHO eigenvalues of Eq.(\ref{Supp:QHO}), given in Eq.(\ref{Supp:eigenQHO}), are shown in red lines as a reference. The anharmonicity, i.e. the unavoidable discrepancy between the Transmon and the QHO model, is noticeable and increases with the eigenvalue order.

\newpage
\section{Classic propagation of a $1D$ wave in a finite periodic media}
\label{notinflong}

\subsection{Spectral analysis}

We now consider the problem of Eq.(\ref{Supp:Helmholtz}) but with a finite length $L$ instead of the infinitely long scenario we implicitly assumed in Section S\ref{inflong}. We recall the equation we are studying is the generic d'Alembert equation
 \begin{equation}
\label{Supp:wavedalembert}
\frac{\partial^2 u(x,t)}{\partial x^2} - F(x)\frac{\partial^2 u(x,t)}{\partial t^2} = 0 \qquad \text{with $u(0,t) = u(L,t) = 0$}
\end{equation}
where $F(x)$ is a $\lambda$-periodic function with dimension of $1/c^2$ and the wave $u(x,t)$, being dimensionless since no properties have been defined, is now constrained in $x=0$ and $x=L$ by the Dirichlet boundary conditions $u(0,t) = 0$ and $u(L,t) = 0$ at any time $t$. To avoid complex boundary effects, we restrict ourselves to a box of size $L$ with a finite number $P$ of wavelengths $\lambda$ so that
 \begin{equation}
\label{Supp:finitewavelengths}
L=P\lambda \qquad \text{with $P \in \mathbb{N}^+$.}
\end{equation}
To continue with the classic harmonically varying medium already considered in Section S\ref{inflong}, we choose $F(x)$ in the form
 \begin{equation}
\label{Supp:periodmedia}
F(x)=F(x+\lambda)=C_0-C_1\cos(2\pi x/\lambda)
\end{equation}
where $x \in [0 \; P\lambda]$. This $F(x)$ is the same as in Eq.(\ref{Supp:Harmonicmod}) but shifted by a phase of $180$ degrees (we will see the influence of this phase in the next few paragraphs). This $F(x)$ is different from the manuscript where we have taken the $F(x)$ analogous to the QHO problem in the narrow-pass-band limit. The reasoning that follows is general for any spatially periodic function whether it is the one here or the one in the manuscript.


Eq.(\ref{Supp:wavedalembert}) being a linear Partial Differential Equation (PDE) with a $\lambda$-periodic coefficient $F(x)$, we know from Floquet-Bloch theory we can look for stationary solutions in the form $u(x,t)=\phi(x)q(t)$ where the spatial part verifies $\phi(x)=\phi(x+\lambda)$ and $q(t) \propto e^{-i \omega t}$ represents a time-harmonic wave oscillating at frequency $\omega$. Replacing this expression in Eq.(\ref{Supp:wavedalembert}), one can separate time and space  and obtain two independent equations:
\begin{equation}
\left\{
\begin{split}
\phi''(x)+\omega^2F(x)\phi(x)=0 & \quad \text{with $\phi(0)=\phi(L)$=0,} \\
\ddot{q}(t)+\omega^2q(t)=0 & \quad \text{with $q(0)=q_0$ and $\dot{q}(0)=v_0$}\\
\end{split}
\right.
\label{Supp:linperiod}
\end{equation}
where $\dot{(\,)}$ and $(\,)'$ denote derivative with respect to time and space, respectively, and $q_0$ and $v_0$ are the initial conditions in modal space. Like in the manuscript, Eq.(\ref{Supp:linperiod}) is linear and the idea is to compute an orthonormal basis for $u(x,t)$ in the form $u(x,t)=\sum_m^{\infty}q_m(t)\phi_m(x)$ with $\phi_m(x)$ solution of the spatial part of Eq.(\ref{Supp:linperiod}) and $q_m(t)=q_0\cos(\omega_m t)+(v_0/\omega_m)\sin(\omega_m t)$ verifying the second part of Eq.(\ref{Supp:linperiod}). Because Eq.(\ref{Supp:linperiod}) is a classical linear wave equations, $\phi_m(x)$ and $\omega_m$ are the eigenfunctions and eigenvalues of an eigenvalue problem, here weighted by the function $F(x)$, that reads
 \begin{equation}
\label{Supp:finiteoperator}
-\frac{d^2 \phi_m(x)}{dx^2}=F(x)\omega_m^2\phi_m(x) \quad \text{with $\phi_m(0)=\phi_m(L)=0$ $\forall m$.}
\end{equation}
In this section, $F(x)$ is given by Eq.(\ref{Supp:periodmedia}) with a harmonically varying $F(x)$ slightly varying about a positive value $C_0=1/c^2$, when the manuscript focused on cases where $F(x)$ is almost everywhere negative, a.k.a. the narrow-pass-band regime.

\begin{figure}[!t]
\centering
	\includegraphics[width=1\columnwidth]{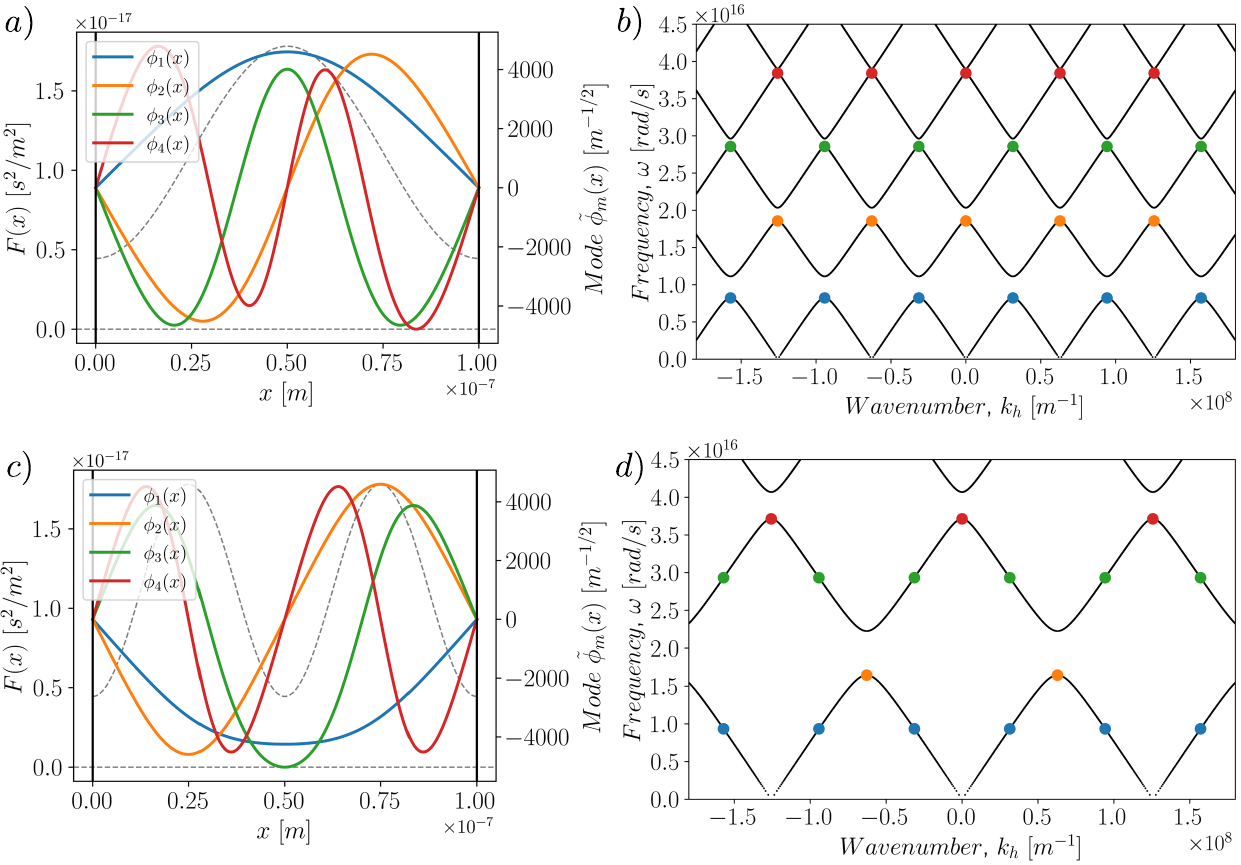}
	\caption{Numerical spectrum of Eq.(\ref{Supp:finiteoperator}) for $C_0=1/c^2$, $C_1=0.6C_0$ and $L=1 \times 10^{-7}$ m. a) Evolution of the function $F(x)$ and the first four normalized stationary modes $\tilde{\phi}_m(x)$ ($\int_0^L\tilde{\phi}^2_m(x)dx=1$) when the medium is made of one wavelength such that $\lambda = L$. b)  Evolution of the first four associated oscillation frequencies $\omega_m$ as a function of discrete extended wavenumber spectrum $k_m\pm h\pi/ L$ with $h \in \mathbb{N}^+$. The continuous dispersion relation corresponding to the infinitely long problem of Figure \ref{Supp:Figure1_classicdispersionrelation}b is shown in thin black lines. c) Evolution of the function $F(x)$ and the first four normalized stationary modes $\tilde{\phi}_m(x)$ when the medium is made of two wavelengths such that $\lambda = L/2$. b)  Evolution of the first four oscillation frequencies $\omega_m$ as a function of discrete extended wavenumber spectrum $k_m \pm h\pi/ (L/2)$ with $h \in \mathbb{N}_0^+$. Thin black lines show the continuous dispersion relation corresponding to the infinitely long problem.}
\label{Supp:Figureannex2_classicwavesinabox}
\end{figure}

When the box length is an integer multiple of the spatial period,
$L=P\lambda$ with $P\in\mathbb{N}^+$, the finite-domain eigenvalue
problem Eq.~(\ref{Supp:finiteoperator}) can be interpreted as a restriction
of the infinite Bloch problem described in Section S\ref{inflong}.
For a frequency within a pass band, the infinite periodic
medium admits two Bloch solutions of the form
$\Psi(x)e^{ikx}$ and its complex conjugate
$\overline{\Psi}(x)e^{-ikx}$, where $\Psi(x+\lambda)=\Psi(x)$.
On the finite interval $[0,L]$, admissible eigenfunctions $\phi(x)$ are
linear combinations of these two counter-propagating waves.

Imposing the Dirichlet boundary conditions $\phi(0)=\phi(L)=0$
selects only specific values of the fundamental wavenumber $k$.
Since $L=P\lambda$ and $\Psi(x)$ is $\lambda$-periodic,
the boundary conditions restrict the Bloch phase $e^{ikL}$,
so that the continuous Bloch parameter becomes discrete:
\begin{equation}
k_\ell=\frac{\ell\pi}{L}=\frac{\ell\pi}{P\lambda},
\qquad \ell=1,\dots,P.
\label{Supp:discretek}
\end{equation}

Let $\omega_n(k)$ denote the dispersion relation of the $n$-th Bloch
band of the infinite periodic medium, $n=0,1,2,\dots$.
Evaluating these dispersion branches at the discrete wavenumbers
$k_\ell$ defined in Eq.~(\ref{Supp:discretek}), modulo $2\pi/\lambda$,
yields the eigenfrequencies of the finite-domain problem Eq.~(\ref{Supp:finiteoperator}).
More precisely, for each pass band $n$,
exactly $P$ eigenfrequencies are obtained in the reduced first Brillouin zone
$[0,\pi/\lambda]$, which read, for $\ell=1,\dots,P$:
\begin{equation}
\omega_{n,\ell} = \omega_n(k_\ell) \text{ for even $n$} 
\qquad 
\omega_{n,\ell} = \omega_n(k_{\ell-1}) \text{ for odd $n$.}
\label{Supp:discretewavi}
\end{equation}
Therefore, the spectrum of Eq.~(\ref{Supp:finiteoperator}) consists of the
collection of $\omega_{n,\ell}$ which, when ordered increasingly, produces the
sequence of eigenfrequencies $\{\omega_m\}_{m\ge1}$ of the finite-domain
problem. From a Floquet perspective, this means that instead of a continuum of Floquet multipliers $\rho$ with $|\rho|=1$ within a given pass band of the infinite Bloch problem, the finite system admits only $P$ such multipliers ($P$ equally spaced complex numbers on the unit circle). This discreteness is a classical consequence of imposing finite-domain boundary conditions on waves, notably in periodic media.

Note that the particular eigenfrequencies corresponding to
$k_\ell=\pi/\lambda$, located at the boundary of the first
Brillouin zone $(-\pi/\lambda,\pi/\lambda]$ for $\ell = P$, require special care.
At this edge, the Floquet multiplier $\rho_{\ell}=e^{ik_{\ell}L}$ equals $-1$, and the associated
standing-wave solutions depend on the spatial phase of the periodic coefficient $F(x)$. Consequently, the
corresponding eigenfrequencies $\omega_{n,\ell}=\omega_n(\pi/\lambda)$
may depend on this phase convention.

\begin{figure}[!b]
\centering
	\includegraphics[width=1\columnwidth]{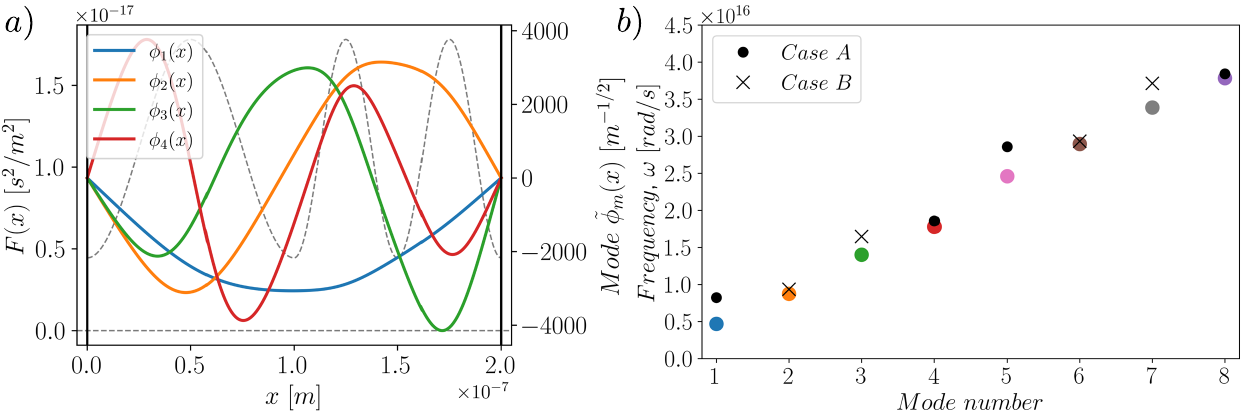}
	\caption{Numerical spectrum of Eq.(\ref{Supp:finiteoperator}) in the case of the concatenation of the two $F(x)$ shown in Figures~\ref{Supp:Figureannex2_classicwavesinabox}a (that we denote $F_A(x)$) and \ref{Supp:Figureannex2_classicwavesinabox}c ($F_B(x)$). a) Evolution of the concatenated function $F(x) = [F_A(x) \; F_B(x)]$ and the first four normalized stationary modes $\tilde{\phi}_m(x)$ such that $\int_0^L\phi^2_m(x)dx=1$. b) Evolution of the first eighth associated oscillation frequencies $\omega_m$ as a function of mode number $m$. The frequencies associated with $F_A(x)$ shown in Figure\ref{Supp:Figureannex2_classicwavesinabox}b are reported with black disks. The frequencies associated with $F_B(x)$ shown in Figure\ref{Supp:Figureannex2_classicwavesinabox}d are reported with black crosses.}
\label{Supp:Floquetstyleannex}
\end{figure}

Figure \ref{Supp:Figureannex2_classicwavesinabox} shows the first four $(\phi_m(x),\omega_m)$ of Eq.(\ref{Supp:finiteoperator}) for $C_0=1/c^2$, $C_1=0.6C_0$ and $L=1\times 10^{-7}$ m. The dimensionless modes $\tilde{\phi}_m(x)$ are normalized by $\sqrt{\lambda/\pi}$ so that $\int_0^L\tilde{\phi}_m(x)^2dx=1$ has no dimension and $\tilde{\phi}_m(x)=\phi_m(x)\sqrt{\pi/\lambda}$ has dimension $[m^{-1/2}]$. When only $P=1$ wavelength $\lambda$ is contained in the box as shown in Figure \ref{Supp:Figureannex2_classicwavesinabox}a-b, only one eigenfrequency $\omega_m=\omega_{n,\ell}$ and associated mode $\tilde{\phi}_m(x)=\tilde{\phi}_{n,\ell}(x)$ are selected among the pass bands of the corresponding infinitely long problem shown in Figure~\ref{Supp:Figure1_classicdispersionrelation}. Note that given the phase of the cosine we have chosen for $F(x)$ given in Eq.(\ref{Supp:periodmedia}) and illustrated in Figure~\ref{Supp:Figureannex2_classicwavesinabox}a, the frequencies $\omega_m$ correspond to the local maxima of the extended dispersion relations in each pass band of the infinitely long problem. Those frequencies can move in the forbidden bands of the infinitely long problem if the phase of $F(x)$ is changed relatively to the boundaries of the box.

In the case of a box containing  $P>1$ wavelengths as in Figure~\ref{Supp:Figureannex2_classicwavesinabox}c-d for $P=2$, there are now $P$ modes and eigenfrequencies per propagating band of the infinite problem. For $P=2$, the modes $\tilde{\phi}_m(x)$ and associated frequencies $\omega_m$ are qualitatively similar to their $P=1$ counterparts plotted in Figure \ref{Supp:Figureannex2_classicwavesinabox}a-b. Contrary to the manuscript, where we place ourselves in the narrow-pass-band limit, it is usually not possible to get an explicit analytical solution for $\tilde{\phi}_m(x)=\tilde{\phi}_{n,\ell}(x)$ and $\omega_m=\omega_{n,\ell}$ when the modulation $C_1$ is small but finite and the modulation wavelength is of the same order of magnitude as the wavelength of the modes (we can not use perturbation or homogenization methods for example). In the case of Figure~\ref{Supp:Figureannex2_classicwavesinabox}, only numerical simulations allow us to describe the dispersion relation of the infinite problem, hence the discrete spectrum of the corresponding finite problem. The narrow-pass-band regime is therefore remarkable since it is a new example of waves in a strongly varying medium where analytical approaches are possible.

Figure~\ref{Supp:Floquetstyleannex} shows the numerical spectrum of Eq.(\ref{Supp:finiteoperator}) for a box made of a medium characterized by a $F(x)$ concatenated with the two functions $F_A(x)$ and $F_B(x)$ illustrated in Figures~\ref{Supp:Figureannex2_classicwavesinabox}a and \ref{Supp:Figureannex2_classicwavesinabox}c. Because the modes $\tilde{\phi}_m(x)$ are not localized in the periodic cells as in the narrow-pass-band limit discussed in the manuscript, the discrete numerical spectrum of $F(x)=[F_A(x) \; F_B(x)]$ is not the union of the spectra of each sub-cell. Like with classical metamaterials in photonic, the medium characterized by $F(x)=[F_A(x) \; F_B(x)]$ is a new medium whose spectrum differs slightly from the one computed in Figures~\ref{Supp:Figureannex2_classicwavesinabox} and it is therefore necessary to numerically compute again Eq.(\ref{Supp:finiteoperator}).

\subsection{Energy of the stationary waves}

In this subsection, we now study the total energy of the stationary waves contained in the box. We compute the total energy of the dimensionless wave $u(x,t)$ solution of Eq.(\ref{Supp:wavedalembert}) with given initial conditions $u(x,0)=u_0(x)$ and $\partial u(x,0)/\partial t=v_0(x)$. The starting point of the energy derivation is to write the spatial part of Eq.(\ref{Supp:linperiod}) or the eigenproblem Eq.(\ref{Supp:finiteoperator}) in weak form, that is
\begin{equation}
\omega_m^2\int_0^LF(x)\tilde{\phi}_m(x)\tilde{\phi}_n(x)dx=-\int_0^L\tilde{\phi}''_m(x)\tilde{\phi}_n(x)dx
\label{Supp:weakform}
\end{equation}
where each equilibrium equation in $m$ has been multiplied by $\tilde{\phi}_n(x)$ and integrated between $0$ and $L$. Using integration by parts and taking into account the Dirichlet Boundary conditions $\tilde{\phi}_m(0)=\tilde{\phi}_m(L)=0$ for all $m$, we get
\begin{equation}
\omega_m^2\int_0^L F(x)\tilde{\phi}_m(x)\tilde{\phi}_n(x)\,dx
= \int_0^L \tilde{\phi}'_m(x)\tilde{\phi}'_n(x)\,dx
   - \left[\tilde{\phi}'_m(x)\tilde{\phi}_n(x)\right]_0^L =\int_0^L \tilde{\phi}'_m(x)\tilde{\phi}'_n(x)\,dx.
\label{Supp:weakformipp}
\end{equation}
Taking the difference between Eq.(\ref{Supp:weakformipp}) evaluated at $m$ and at $n$, one can write the classic orthogonal equation
\begin{equation}
(\omega_m^2-\omega_n^2)\int_0^L F(x)\tilde{\phi}_m(x)\tilde{\phi}_n(x)\,dx = 0.
\label{Supp:ortho}
\end{equation}
If $m\neq n$, $(\omega_m^2-\omega_n^2)\neq 0$ and we obtain the first orthogonality relation that is 
\begin{equation}
\int_0^L F(x)\tilde{\phi}_m(x)\tilde{\phi}_n(x)\,dx = 0 \qquad \text{if $m\neq n$.}
\label{Supp:ortho1}
\end{equation}
Eq.(\ref{Supp:ortho1}) is the classic weighted orthogonality condition for linear waves in periodic media that holds in the narrow-pass-band limit (because of the near numerical degeneracy in this limit, the numerical eigenvalue solver could need particular treatment such as Gram-Schmidt procedure or generalized symmetric solvers). Replacing Eq.(\ref{Supp:ortho1}) for $m \neq n$ in Eq.(\ref{Supp:weakformipp}), we find the second orthogonality relation
\begin{equation}
\int_0^L \tilde{\phi}'_m(x)\tilde{\phi}'_n(x)\,dx = 0 \qquad \text{if $m\neq n$.}
\label{Supp:ortho2}
\end{equation}
\begin{figure}[!b]
\centering
	\includegraphics[width=1\columnwidth]{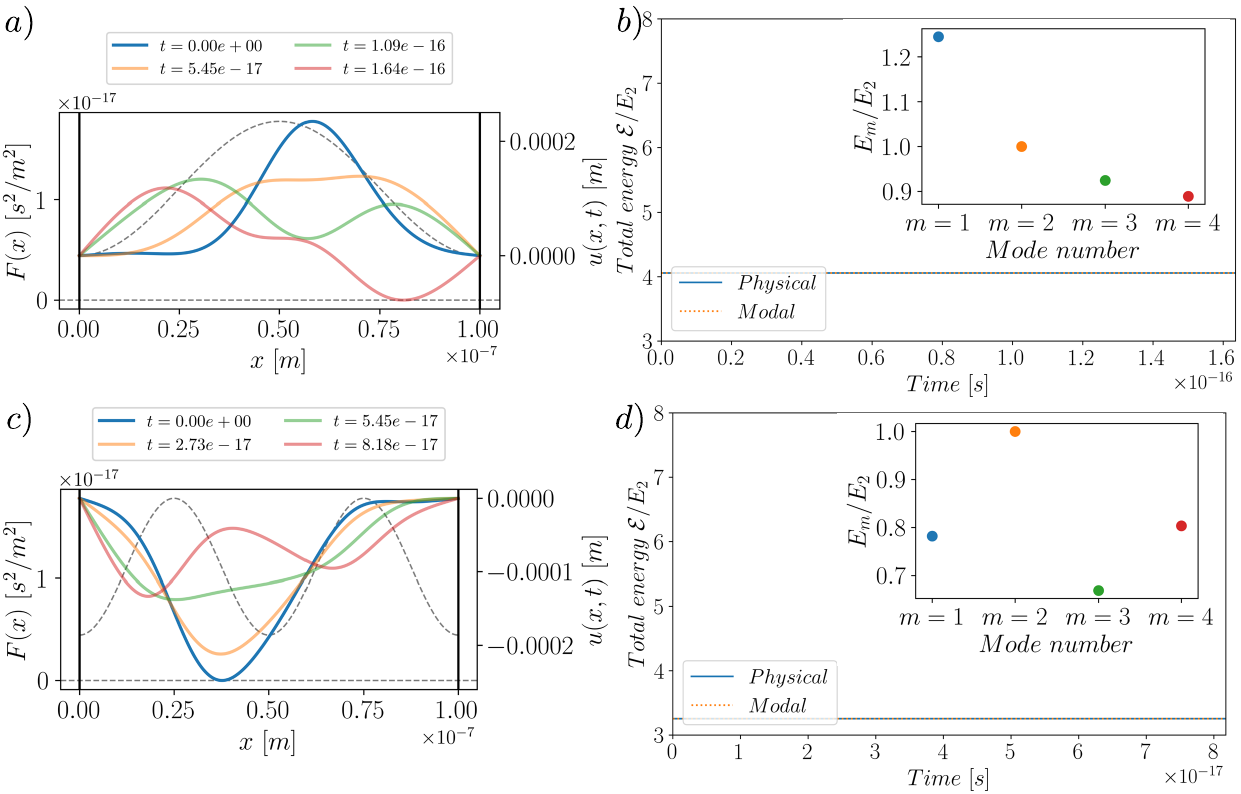}
	\caption{Energy contained in the boxes whose configuration is shown in Figure \ref{Supp:Figureannex2_classicwavesinabox}. a) Four time steps showing the evolution $u(x,t)=\sum_{m=1}^4 \tilde{\phi}_m(x)\tilde{q}_m(t)$ in the medium $F_A(x)$ for $\tilde{q}_m(0)=1/\omega_m$ and $\dot{\tilde{q}}_m(0)=0$. The four modes $\phi_m(x)$ and frequencies $\omega_m$ are shown in Figure \ref{Supp:Figureannex2_classicwavesinabox}a-b. b) Total energy $\tilde{\mathcal{E}}$ in the box as a function of time computed in the modal (Eq.(\ref{Supp:energymodal})) and physical space (Eq.(\ref{Supp:energy_physical2})). Inset shows the energy contribution $\tilde{\mathcal{H}}_m$ for each mode $m$. The energies are normalized by $\tilde{\mathcal{H}}_2$. c) Same as a) but for the medium $F_B(x)$ and the four modes shown in Figures \ref{Supp:Figureannex2_classicwavesinabox}c-d. d) Energies of the wave shown in c).}
\label{Supp:Figureannex2_classicwavesinabox_energy}
\end{figure}
From now, taking the orthogonality relations into account in Eq.(\ref{Supp:weakformipp}), one finds the weak form expression of the frequency of each mode $\omega_m$ 
\begin{equation}
\omega_m^2 = \frac{\int_0^L \tilde{\phi}'_m(x)^2\,dx}{\int_0^L F(x)\tilde{\phi}_m(x)^2\,dx} = \frac{\tilde{k}_m}{\tilde{m}_m}
\label{Supp:weakformfrequency}
\end{equation}
where $\tilde{k}_m$ and $\tilde{m}_m$ are respectively the effective stiffness and mass of the $m^{th}$ mode of vibration $u_m(x,t)=\tilde{\phi}_m(x)\tilde{q}_m(t)$. Thus, projecting the d'Alembert equation Eq.(\ref{Supp:wavedalembert}) on its orthogonal modal basis $\phi_m(x)$, each stationary modes $u_m(x,t)$ oscillating at $\omega_m$ verifies
\begin{equation}
\tilde{m}_m\ddot{\tilde{q}}_m(t)+\tilde{k}_m\tilde{q}_m(t)=0
\label{Supp:harmonicmodes}
\end{equation}
that is the equation of motion of a classic mass-spring system. Because each oscillator is independent in Eq.(\ref{Supp:harmonicmodes}), the total energy inside the box of length $L$ reads, in modal coordinates
\begin{equation}
\tilde{\mathcal{E}}=\sum_m\left(\frac{1}{2}\tilde{k}_m\tilde{q}^2_m(t) + \frac{1}{2}\tilde{m}_m\dot{\tilde{q}}^2_m(t) \right) = \sum_m \tilde{\mathcal{H}}_m
\label{Supp:energymodal}
\end{equation}
where $\tilde{\mathcal{H}}_m$ is the so-called Hamiltonian of the $m^{th}$ mode of oscillations. Note that in this section, because we started with a generic wave equation Eq.(\ref{Supp:wavedalembert}) with no given properties, the energies $\tilde{\mathcal{E}}$ and $\tilde{\mathcal{H}}_m$ in Eq.(\ref{Supp:energymodal}) are homogeneous to a curvature in $m^{-1}$ and not to an actual energy in Joule. To retrieve an energy, one would need to multiply by a parameter in $N.m^2$ such as the bending stiffness for transverse waves in elastic beams or $A/\mu_0$ for electromagnetic waves like in the manuscript. Using the modal projection $u(x,t)=\sum_m\tilde{\phi}_m(x)\tilde{q}_m(t)$ and the definition of $\tilde{k}_m$ and $\tilde{m}_m$ given in Eq.(\ref{Supp:weakformfrequency}), we can express the total energy given in Eq.(\ref{Supp:energymodal}) in the physical space 
\begin{align}
\tilde{\mathcal{E}}
= & \sum_m\frac{1}{2}\left(\tilde{q}_m^2(t)\int_0^L \tilde{\phi}_m'(x)^2\,dx  
      + \dot{\tilde{q}}_m^2(t)\int_0^L F(x)\tilde{\phi}_m(x)^2\,dx \right)
\notag
\\
= & \frac{1}{2}\left[ \int_0^L \left(\frac{\partial u}{\partial x}\right)^2 dx
     + \int_0^L F(x)\left(\frac{\partial u}{\partial t}\right)^2 dx \right]
\label{Supp:energy_physical2}
\end{align}
where we have used the orthogonality conditions in Eq.(\ref{Supp:ortho1}) and Eq.(\ref{Supp:ortho2}) to go from Eq.(\ref{Supp:energymodal}) to Eq.(\ref{Supp:energy_physical2}). From now, we pick $t=0$ to compute the total energy that should be constant over time since the system we study is conservative. To ensure equipartition of energy for each mode, we impose the classic modal initial conditions for each modal coordinate $q_m(t)$ of Eq.(\ref{Supp:harmonicmodes})
\begin{equation}
\tilde{q}_m(0) = \tilde{Q}_0/\omega_m \qquad \text{and} \qquad \dot{\tilde{q}}_m(0) = \tilde{V}_0 = 0 \qquad \text{for all $m$}
\label{Supp:inicond}
\end{equation}
where $\tilde{Q}_0$ and $\tilde{V}_0$ are two constants. Taking the initial conditions Eq.(\ref{Supp:inicond}) in the expression of total energy in modal coordinates given in Eq.(\ref{Supp:energymodal}), we get 
\begin{equation}
\tilde{\mathcal{E}}=\sum_m\frac{1}{2}\tilde{m}_m\left(\tilde{Q}_0^2+\tilde{V}_0^2\right)=\sum_m\frac{\tilde{Q}_0^2+\tilde{V}_0^2}{2}\int_0^LF(x)\tilde{\phi}_m(x)^2dx=\sum_m\tilde{\mathcal{H}}_m
\label{Supp:energymodalt0}
\end{equation}

We see from Eq.(\ref{Supp:energymodalt0}) that in the particular case where $F(x)$ is a constant, the normalization condition $\int_0^L\tilde{\phi}_m(x)^2dx=1$ ensures equipartition of energy $\tilde{\mathcal{H}}_m$ between each mode. We have shown in the manuscript that in the narrow-pass-band limit, analytical expressions are attainable thanks to the quantum mathematical analogy and  we recover the equipartition of energy. But in the typical case where $F(x)=F(x+\lambda)$ periodically varies about a positive value, we will see $\int_0^LF(x)\tilde{\phi}_m(x)^2dx$ is usually no more a constant and the energy varies from one mode to another. Furthermore, unless the periodic variation is small or ``fast'', in which case one could use perturbation methods, it is often too involved to analytically estimate the total energy.

\begin{figure}[!b]
\centering
	\includegraphics[width=1\columnwidth]{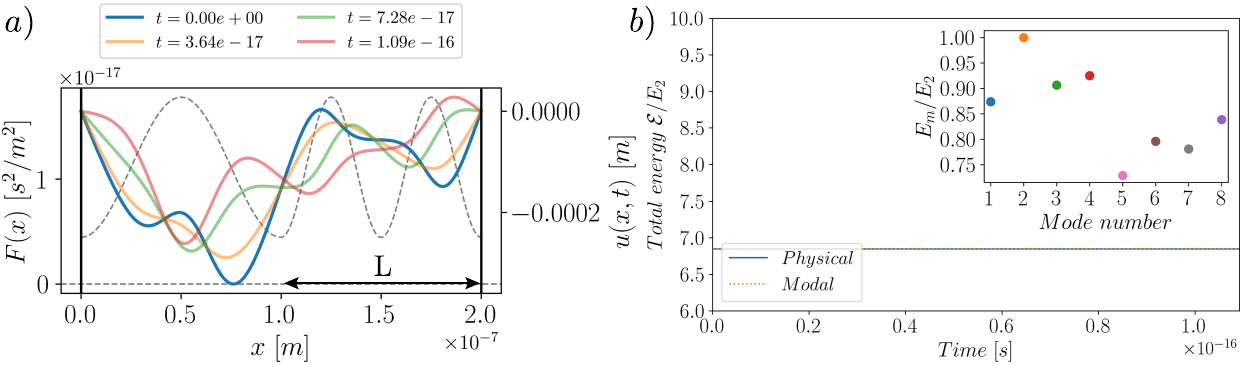}
	\caption{Energy contained in the box whose configuration is shown in Figure \ref{Supp:Floquetstyleannex}. a) Four time steps showing the evolution $u(x,t)=\sum_{m=1}^8 \tilde{\phi}_m(x)\tilde{q}_m(t)$ in the concatenated medium $F(x) = [F_A(x) \; F_B(x)]$ for $\tilde{q}_m(0)=1/\omega_m$ and $\dot{\tilde{q}}_m(0)=0$. The eight modes $\phi_m(x)$ and frequencies $\omega_m$ are shown in Figure \ref{Supp:Floquetstyleannex}. b) Total energy $\tilde{\mathcal{E}}$ in the box as a function of time computed in the modal (Eq.(\ref{Supp:energymodal})) and physical space (Eq.(\ref{Supp:energy_physical2})). Inset shows the energy contribution $\tilde{\mathcal{H}}_m$ for each mode $m$. The energies are normalized by $\tilde{\mathcal{H}}_2$.}
\label{Supp:Figureannex2_classicwavesinabox_concatenate_energy}
\end{figure}

Figures \ref{Supp:Figureannex2_classicwavesinabox_energy}a and \ref{Supp:Figureannex2_classicwavesinabox_energy}c show four time steps of the evolution $u(x,t)=\sum_1^4 \tilde{\phi}_m(x)\tilde{q}_m(t)$ for the configuration and a modal basis shown in Figure \ref{Supp:Figureannex2_classicwavesinabox}a-b and \ref{Supp:Figureannex2_classicwavesinabox}c-d, respectively. The initial conditions are chosen to be $\tilde{q}_m(0)=\tilde{Q}_0/\omega_m$ with $\tilde{Q}_0=2$ $m^{1/2}s^{-1}$ and $\dot{\tilde{q}}_m(0)=\tilde{V}_0=0$ for the $N=4$ modes in both cases. The total energy $\tilde{\mathcal{E}}$ of the stationary waves $u(x,t)$, normalized by the energy of their second mode $\tilde{\mathcal{H}}_2$, is given in Figures \ref{Supp:Figureannex2_classicwavesinabox_energy}b and \ref{Supp:Figureannex2_classicwavesinabox_energy}d. As expected, the computation in modal or physical space, through Eq.(\ref{Supp:energymodal}) or Eq.(\ref{Supp:energy_physical2}), respectively, lead to the same result that is: i) a total energy conserved with time and ii) a total energy that is the sum of the contribution of each mode $\tilde{\mathcal{H}}_m$ as shown by the inset of Figures \ref{Supp:Figureannex2_classicwavesinabox_energy}b and \ref{Supp:Figureannex2_classicwavesinabox_energy}d. Unlike the narrow-pass-band limit where the modal energies $\tilde{\mathcal{H}}_m$ were all equal for a fixed amplitude of $F(x)$, there is here no equipartition of energy between the modes for a wave oscillating in a box made of a medium with slightly modulated positive properties. Indeed, apart in the very particular case of the quantum analogy uncovered in the manuscript, there is no reason for the integral $\int_0^LF(x)\tilde{\phi}_m(x)^2dx$ to be constant for different mode number $m$.

Figure \ref{Supp:Figureannex2_classicwavesinabox_concatenate_energy}a shows four time steps of the evolution $u(x,t)=\sum_1^8 \tilde{\phi}_m(x)\tilde{q}_m(t)$ for the concatenated medium characterized by $F(x)=[F_A(x) \; F_B(x)]$ and a modal basis shown in Figure \ref{Supp:Floquetstyleannex}. The initial conditions are chosen to be $\tilde{q}_m(0)=\tilde{Q}_0/\omega_m$ with $\tilde{Q}_0=2$ $m^{1/2}s^{-1}$ and $\dot{\tilde{q}}_m(0)=\tilde{V}_0=0$ for the $N=8$ modes. The total conserved energy $\tilde{\mathcal{E}}$ of the stationary wave $u(x,t)$, normalized by the energy of the second mode $\tilde{\mathcal{H}}_2$, is given in Figure \ref{Supp:Figureannex2_classicwavesinabox_concatenate_energy}b when its modal projection is given in inset.

Since there is no localization properties like in the narrow-pass-band limit, the spectrum of the box made of a concatenated $F(x)=[F_A(x) \; F_B(x)]$ are not simply the union of the spectra of each independent sub-medium, and the same is true for the modal energies $\tilde{\mathcal{H}}_m$. The medium characterized by $F(x)=[F_A(x) \; F_B(x)]$ is a whole new one and the $\tilde{\mathcal{H}}_m$ could not be deduced from Figures \ref{Supp:Figureannex2_classicwavesinabox_energy}b, d. Again, in the classic framework of waves propagating in strongly varying continuum media, there are usually no simple rules or analytical expressions that can be derived and one has no choice but to numerically compute the energy of the waves oscillating in the box.


\end{document}